\RequirePackage{fix-cm}
\documentclass[smallextended]{svjour3}       % onecolumn (second format)
\smartqed  % flush right qed marks, e.g. at end of proof
\usepackage{graphicx}
\usepackage{colortbl}
\usepackage{comment}
\usepackage{authblk}
\makeatletter
\newcommand*{\rom}[1]{\expandafter\@slowromancap\romannumeral #1@}
\makeatother
\usepackage{cite}

\usepackage{color}
\usepackage{graphicx}

\usepackage{multirow}
\usepackage{amsmath}
\usepackage{algorithmic}
\usepackage[ruled,norelsize]{algorithm2e} % For algorithms

\SetAlFnt{\small}
\SetAlCapFnt{\small}
\SetAlCapNameFnt{\small}
\SetAlCapHSkip{0pt}
\IncMargin{-\parindent}

\usepackage{mathtools}
\DeclarePairedDelimiter{\ceil}{\lceil}{\rceil}

\begin{document}

\title{Data-Driven Stochastic Optimization for Power Grids Scheduling under High Wind Penetration
}

\titlerunning{Data-Driven SUC for Power Grid Scheduling with Renewable Energy}        % if too long for running head

\author{Wei Xie  \and
Yuan Yi         \and
        Zhi Zhou  \and
        Keqi Wang
}


\institute{
           Corresponding author: Wei Xie (w.xie@northeastern.edu) \at
            Northeastern University, Boston, MA 02115 \\
            \and
            Yuan Yi \at
              Rensselaer Polytechnic Institute, Troy, NY 12180 \\
 %             Tel.: +123-45-678910\\
 %             Fax: +123-45-678910\\
%              \email{fauthor@example.com}           %  \\
           \and
           Zhi Zhou \at
            Argonne National Laboratory, Argonne, IL 60439 \\
            \and
           Keqi Wang \at
            Northeastern University, Boston, MA 02115 \\
}

\date{Received: date / Accepted: date}

\maketitle
%\vspace{-in}

\begin{abstract}
		To address the environmental concern and improve the economic efficiency, the wind power is rapidly integrated into smart grids. However, the inherent uncertainty of wind energy raises operational challenges. To ensure the cost-efficient, reliable and robust operation, it is critically important to find the optimal decision that can correctly and rigorously hedge against all sources of uncertainty. In this paper, we propose data-driven stochastic unit commitment (SUC) to guide the power grids scheduling. Specifically, given the finite historical data, the posterior predictive distribution is developed to quantify the wind power prediction uncertainty accounting for both inherent stochastic uncertainty of wind power generation and input model estimation error.
		For complex power grid systems, a finite number of scenarios is used to estimate the expected cost in the planning horizon. 
		To further control the impact of finite sampling error induced by using the sample average approximation (SAA), we propose a parallel computing based optimization solution methodology, which can quickly find the reliable optimal unit commitment decision hedging against various sources of uncertainty. The empirical study over six-bus and 118-bus systems demonstrates that our approach can provide more efficient and robust performance than the existing deterministic and stochastic unit commitment approaches. 

\end{abstract}

\begin{comment}

\section*{Declarations}

\noindent {\textbf{Funding:}}
This research received no specific grant from any funding agency in the public, commercial, or not-for-profit sectors.

\noindent {\textbf{Conflicts of interest:}}
The authors declare that there is no conflict of interests regarding the publication of this article.

\noindent {\textbf{Availability of data and material:}}
Due to confidentiality agreements, the wind power generation data are not publicly available.

\noindent {\textbf{Code availability:}}
All code of proposed data-driven SUC framework is publicly available at https://github.com/kw48792/data-driven-suc.

\noindent {\textbf{Authors' contributions:}}
Wei Xie, Yuan Yi and Zhi Zhou initiated the study. Wei Xie, Yuan Yi and Keqi Wang proposed data-driven SUC approach, developed the optimization algorithm, and conducted the experiments. Zhi Zhou offered the power grid unit commitment problem knowledge support, assisted with the methodology development, provided the data about power system and wind power generation, and assessed the performance of proposed framework. Wei Xie wrote the paper in conjunction with Zhi Zhou, Yuan Yi and Keqi Wang. All authors read and approved the final manuscript, contributing edits. Wei Xie takes full responsibility for the work, including the study design, the decision to submit and publish the manuscript.

\end{comment}

\section{INTRODUCTION}
	\label{intro}
	Wind power is rapidly incorporated into power grids in an effort to combat the climate change and improve power system resilience \cite{Hargreaves_Hobbs_2012, Zhou_etal_2013, Zhao_Wu_2014}. In the past few years, the wind energy capacity expanded explosively \cite{Jiang_Wang_Guan_2012, Hargreaves_Hobbs_2012, Zhao_Wu_2014}. It is also projected that the wind power penetration will continue to grow in the near future \cite{Zhou_etal_2013}. 
	However, the inherent volatility of wind energy has a significant impact on the system operation
	\cite{Jiang_Wang_Guan_2012, Hargreaves_Hobbs_2012, Papavasiliou_Oren_2013}. To ensure a cost-efficient and reliable power grid scheduling, the stochastic unit commitment (SUC) model is widely used in the literature, especially under the situations with high wind penetration \cite{Wang_Shahidehpour_Li_2008, Ruiz_etal_2009, Tuohy_etal_2009, Wang_Guan_Wang_2012}. Decision makers seek the unit commitment decision minimizing the total \textit{expected} cost of the power production to meet the demand, which explicitly accounts for the inherent stochastic uncertainty of wind power generation \cite{Wang_Shahidehpour_Li_2008, Ruiz_etal_2009, Tuohy_etal_2009, Zhou_etal_2013}.

	However, the existing SUC approaches tend to ignore two sources of uncertainty, which can lead to inferior and unreliable unit commitment decisions. First, the underlying true statistical input model characterizing the wind power generation uncertainty is unknown and estimated by finite historical data, which can induce the input model estimation uncertainty, called \textit{model risk}. The existing SUC approaches, called  \textit{the empirical approach}, tend to take the estimated statistical model as the true one \cite{Ruiz_etal_2009, Tuohy_etal_2009, Wang_Guan_Wang_2012} and ignore \textit{the model estimation error} \cite{Xie_etal_2018}.  
	Second, to solve SUC, the sample average approximation (SAA), using finite number of scenarios to approximate the expected cost in the planning horizon, can introduce \textit{the finite sampling error}, which is also typically ignored in the existing SUC approaches \cite{Ruiz_etal_2009, Tuohy_etal_2009, Wang_Guan_Wang_2012}.

	In this paper, we introduce a data-driven SUC model and further develop a parallel computing based optimization solution methodology. Our study can lead to the optimal unit commitment decision which can appropriately hedge against all sources of uncertainty. Basically, both parametric and nonparametric statistical model can be used to characterize the inherent stochastic uncertainty of wind power generation. Since the underlying input model is unknown and estimated by finite historical data, this induces the model estimation uncertainty and we quantify it with the posterior distribution. Then, we utilize the posterior predictive distribution to quantify the wind power generation prediction uncertainty, which accounts for both stochastic uncertainty and model estimation error. Thus, driven by the scenarios generated by the posterior predictive distribution, we propose the \textit{data-driven SUC}, which leads to the optimal decision simultaneously hedging against both wind power generation stochastic uncertainty and input model estimation error. 
	
	Built on data-driven SUC, we further introduce a parallel computing based optimization approach, called \textit{the optimization and selection (OPSEL)}, which can efficiently control the impact from finite sampling error induced by SAA. 
	Specifically, to solve SUC problems, we observe that the computational effort is heavily invested in searching for the optimal unit commitment decision. Compared with the optimization search, given a candidate unit commitment decision, it takes much less time to assess its performance. %solving a large number of economic dispatch problems only takes a small fraction of time. 
	\textit{Thus, the OPSEL approach includes the parallel optimization search and the best candidate decision selection.}  In the optimization step, we utilize the parallel computing to simultaneously solve a sequence of finite sample approximated data-drive SUC problems and obtain candidate solutions. Then, in the selection step, we use the rank and selection approach to efficiently evaluate these candidate decisions and select the best decision. The proposed approach can be used by general SUC to control the impact of finite sampling error and quickly find the optimal unit commitment decision, which is critically important for guiding the dynamic scheduling decision for complex power grids in the intra-day market.

	The main contributions of this paper are listed as follows.
	\begin{enumerate}
		\item To the authors' best knowledge, there is no existing approach explicitly accounting for all three sources of uncertainty: (1) inherent stochastic uncertainty of wind power generation, (2) SUC input model estimation uncertainty, and (3) finite sampling error induced by using the sample average approximation on the expected cost in the planning horizon of SUC. We propose a data-driven stochastic optimization framework that can appropriately hedge against all sources of uncertainties and quickly deliver a reliable cost-efficient optimal dynamic unit commitment decision.
		
		\item The proposed data-driven SUC leads to the optimal unit commitment decision \textit{simultaneously} hedging against both the inherent stochastic uncertainty of wind power generation and the SUC input model estimation uncertainty. It can be applied to cases with either parametric or nonparametric wind power forecast models.
		
		\item Our OPSEL approach can utilize the parallel computing and quickly solve for the optimal unit commitment decisions hedging against the impact of finite sampling error induced by SAA, which is large especially for complex power grids with high wind power penetration. 	
		
	\end{enumerate}
	
	%\textcolor{red}{(Keqi, we need to present out the key insights of proposed data-driven SUC modeling and optimization solution methodology. And then briefly summarize the empirical study observations to further support them.)}
	
	The organization of this paper is as follows. We review the related literature in the next section. We formally state our problem and introduce the proposed data-driven SUC modeling in Section~\ref{sec:problemDescription}. In Section~\ref{sec:framework}, we introduce the OPSEL approach that can quickly solve SUC and control the impact of SAA finite sampling error.  Both six-bus and 118-bus test cases are used to study the performance of our approach in Section~\ref{sec:emp} and the results demonstrate that the proposed data-driven SUC framework has the clear advantages. We conclude this paper in Section~\ref{sec: conclusion}. The code of proposed data-driven SUC framework is available on GitHub at  https://github.com/kw48792/data-driven-suc.

\section{LITERATURE REVIEW}
\label{sec:litreview}
	
%\textcolor{red}{(Keq, please provide more clear and comprehensive literature review, which makes the contributions of our paper easy to follow. At the end of this section, we can discuss: (1) the proposed framework belongs to which stream of research; and (2) present the key contributions compared to the existing studies)}
	
In the power system literature, several streams of optimization approaches have been developed for the stochastic unit commitment (SUC) problem.  
The first one and the most commonly used one is called \textit{the empirical SUC}. Given the historical data, it first estimates the underlying statistical input model characterizing the wind power generation variation, and then takes the estimate as the true one. 
Stochastic programming was first introduced to solve the unit commitment problem with uncertainty from load \cite{Takriti_etal_1996}. This approach has frequently been applied in recent research, as renewables are integrated into the power system on a large scale; 
%In \cite{madaeni2013impacts}, the authors focus on the effect of wind uncertainty on the scheduling of conventional generators and the resulting cost impacts. They compare the benefits of cost reduction from SUC with another potential solution, called demand response. The results show that the demand response is significantly more effective. Further, the subadditive benefits are reported using the two solutions together. 
see the examples in \cite{Tuohy_etal_2009, Wang_Guan_Wang_2012,madaeni2013impacts,Huang_Zheng_Wang_2014}.
While the empirical SUC accounts for the inherent stochastic uncertainty of renewable energies, it fails to account for statistical input model estimation uncertainty and finite sampling error induced by SAA, which could lead to inferior decisions.
	
%%%%%%%%%robust optimization
The second stream is \textit{the robust optimization} (RO). Without assuming the distribution modeling wind power generation variability, this approach focuses on the worst-case scenario, with the objective minimizing the worst-case cost. The studies in \cite{Jiang_Wang_Guan_2012,Bertsimas_etal_2013} employed RO to smart grids with high wind power penetration, \cite{Zhao_etal_2013} further extended RO to multi-stage cases, and  \cite{Lee_etal_2014} included the transmission line constraints in RO. However, RO is too conservative; see \cite{Xiong_etal_2017}. While some efforts have been made to reduce the conservativeness \cite{Jiang_Wang_Guan_2012, Zhao_Zeng_2012, Lorca_Sun_2015,kazemzadeh2019robust}, since RO only considers the worst-case without taking into consideration of the likelihood of all scenarios, the conservativeness issue persists. 
	
%%%%%%distributionally robust optimization
The third stream, called \textit{the distributionally robust optimization} (DRO), is proposed to overcome the limitation of RO; see for example \cite{Bian_etal_2015,Wang_etal_2016J,Xiong_etal_2017}. The distributional robust unit commitment model minimizes the worst-case expected cost over a set of probability distributions, called an ambiguity set, accounting for the input model estimation uncertainty. DRO is a special case of the composite measure approach where separate measures are used to quantify the input model estimation uncertainty and stochastic uncertainty, and then the composite of these measures is used in the objective \cite{Xiong_etal_2017}. Even although this approach could produce less conservative decisions than RO, it fails to take into account the possibility of distribution candidates being the true one. Hence, the resulting scheduling decision is still too conservative and costly.  
	
%%%%%%%minimax regret
The fourth stream is called  \textit{the minimax regret optimization}. The regret is defined by the objective difference between the current solution without knowing the uncertain parameters and the perfect-information solution. %, i.e., the action we would have taken had we known which scenario would occur.  
Jiang et al. \cite{Jiang_etal_2013} introduced an innovative minimax regret unit commitment model aiming to minimize the maximum regret of the day-ahead decision over all possible realizations of the uncertain wind power generation.  While the minimax regret optimization can deliver less conservative results than RO, like DRO, it also fails to take into account the possibility of distribution candidates being the true one. Hence, it suffers the similar drawback as DRO.
	
%%%%%chance,constrain

Additionally, there is another stream called \textit{the chance-constrained programming}. It was first introduced to model the UC problem with random wind power generation in \cite{ozturk2003stochastic, ozturk2004solution}. %Wind power generation is modeled as a multivariate random variable with mean and standard deviation for different time periods. 
The objective is to satisfy the net load (load minus wind) with a specified high probability level over the entire time horizon while minimizing the operating cost. The original SUC problem is decomposed to a sequence of deterministic versions of the UC problem that converge to the solution of the chance-constrained program. In \cite{wang2011chance}, the SUC problem with uncertain wind power generation is formulated as a chance-constrained two-stage stochastic program. A combined sample average approximation (SAA) algorithm is further developed to solve the problem. %The approach can provide a solution that converges to the optimal one as the number of scenarios increases. %Finally, the model is tested on a six-bus system and a revised 118-bus system. The computational results indicate that increasing the utilization of wind power might increase the total power generation cost. %Pozo et al. \cite{pozo2012chance} present a chance-constrained formulation with an alpha-quantile measure to determine the confidence level of meeting demand under K simultaneous contingencies, also factoring in load and demand uncertainty. CVar and duality theory are used to transform the chance-constrained optimization problem to a mixed-integer linear programming problem.
Similar to the empirical SUC, this method relies on the assumption that underlying input model of uncertain variables can be accurately estimated with historical data.

\indent In this paper, we propose a \textit{data-driven SUC} with the prediction risk accounting for both wind power generation stochastic uncertainty and the input model estimation error. 
%Specifically, given the finite historical data, the compound approach develops a posterior predictive distribution accounting for both stochastic and model estimation uncertainty of wind power. 
To assess the performance of any candidate decision, the sample average approximation (SAA) is used to approximate the expected future cost  \cite{shapiro_2009, Wang_Guan_Wang_2012}. %In SAA, a set of finite number scenarios is generated for approximation, hence inducing the finite estimation error. %To control the finite estimation error, we solve multiple SAA problems with distinct sets of scenarios and select the best candidate.  %we need an efficient approach to efficiently and accurately estimate the objective values of each candidate and thus identify the best one. 
%%%%%%%OPSEL
To hedge against the impact of finite sampling error, we propose the OPSEL approach. This approach is built based on \textit{ranking-and-selection techniques}, which are originally introduced for time-consuming stochastic simulation optimization \cite{Chen_etal_2000, Boesel_Nelson_Kim_2003, Kim_Nelson_2007, Powell_Ryzhov_2012}. Since each simulation run could be time-consuming, given a fixed number of candidate system designs and a limited computational budget, they are statistical comparison approaches developed to efficiently find the best design by sequentially allocating more simulation resources to the promising candidates. Existing ranking and selection approaches include the indifference zone method \cite{Kim_Nelson_2001, Boesel_Nelson_Kim_2003},  the Expected Improvement (EI) methods \cite{Chick_Inoue_2001, Frazier_Powell_Dayanik_2008, Chick_Branke_Schmidt_2010}, and Optimal Computing Budget Allocation (OCBA) \cite{Chen_etal_2000, Chen_etal_2008,xiao2013optimal,xu2015simulation}. 
%OCBA is a popular ranking and selection procedure. %The framework was proposed to enhance the simulation efficiency by intelligently allocating replications to each alternative solution based on mean and variance. %
The objective is to maximize simulation efficiency, expressed as the probability of correct selection (PCS) of the underlying best candidate under limited computing budget. %, which is the probability that the alternatives are selected is truly the best. 
%Intuitively, to maximize the correctness of selection, a larger portion of the computing budget should be allocated to those designs that are critical competitors of the optimal answer. 
We consider OCBA in the our proposed OPSEL procedure because it has several advantages. First, it guarantees the asymptotically correct selection  \cite{Glynn_Juneja_2004}. Second, the convergence rate provided by OCBA is at least as good as other ranking-and-selection methods \cite{Ryzhov_2016}. Third, it demonstrates good finite sampling performance in many studies \cite{Quan_etal_2013}. 
	
    \section{PROBLEM STATEMENT AND DATA-DRIVEN SUC}
    \label{sec:problemDescription}
	
	In this section, we first describe the two-stage stochastic unit commitment (SUC) problem in Section~\ref{subsec:two-stageProblem}. Since the underlying input model, characterizing the inherent stochastic uncertainty of wind power generation, is unknown, it is estimated by using finite historical wind power data. This introduces the input model estimation uncertainty. In Section~\ref{subsec:empiricalSUC}, we review the existing empirical SUC approach which takes the estimated model as the true one and ignores the model estimation uncertainty. We also discuss the existing deterministic unit commitment (UC) model, which ignores the prediction risk induced from both wind power inherent stochastic uncertainty and model risk.
	Then, in Section~\ref{subsec:compounduc}, we propose the data-driven SUC accounting for both sources of uncertainties. Our proposed stochastic unit commitment modeling approach can be applied to situations with both parametric and nonparametric forecast model for wind power. 
	
	\subsection{Stochastic Unit Commitment Model}
	\label{subsec:two-stageProblem}
	Let $\pmb{\xi}$ denote the random wind power generation, and let $F^c$ represent the underlying ``correct" statistical input model for SUC with $\pmb{\xi}\sim F^c$. Here, we consider a general formulation of the two-stage stochastic unit commitment problem  \cite{Zheng_Wang_Liu_2015, Wang_etal_2016}  
	\begin{eqnarray}
	\small
	\min_{\mathbf{u}} G(\mathbf{u}) &\equiv & \mathbf{C}^{su} \mathbf{u} + \mbox{E}_{\pmb{\xi} \sim F^c} \left[ \min_{\mathbf{y}} \mathbf{C}^{fuel} \mathbf{y}(\mathbf{u},\pmb{\xi})\right]  
	\label{eq: genobj} \\
	\mbox{s.t. }  && A\mathbf{u} \leq B  
	\label{eq: gen1} \\
	&& H\mathbf{u}+Q\mathbf{y}(\mathbf{u},\pmb{\xi}) \leq M(\pmb{\xi})
	\label{eq: gen2} 
	\end{eqnarray} 
	where $ \mathbf{C}^{su}$ is the first-stage cost coefficient, consisting of various startup and shutdown costs, and the coefficient $ \mathbf{C}^{fuel}$ represents the fuel cost; see \cite{Zheng_etal_2013, Zheng_Wang_Liu_2015, Hobbs_Rothkopf_O'Neill__Chao_2001}.
	The first-stage unit commitment decision for thermal generators, denoted by $\mathbf{u}$, is made prior to the realization of $\pmb{\xi}$.
	The second-stage economic dispatch decision, denoted by $\mathbf{y}(\mathbf{u},\pmb{\xi})$, is made after the unveiling of $\pmb{\xi}$ and it depends on $\mathbf{u}$.

	Objective~(\ref{eq: genobj}) includes the cost incurred in the first stage and the expected dispatch cost incurred in the planning horizon. The general constraints for the first- and second-stage decisions can be expressed in (\ref{eq: gen1}) and (\ref{eq: gen2}). Denote the optimal unit commitment decision by $\mathbf{u}^\star$ and the optimal objective by $G(\mathbf{u}^\star)\equiv \mathbf{C}^{su}\mathbf{u}^\star + \mbox{E}_{\pmb{\xi} \sim F^c} [ \min_{\mathbf{y}} \mathbf{C}^{fuel} \mathbf{y}(\mathbf{u}^\star,\pmb{\xi})]$.

	%   =================
	
	\subsection{Review of Existing Empirical Stochastic Unit Commitment (SUC) and Deterministic Unit Commitment (UC) Approaches}
	\label{subsec:empiricalSUC}

	In this section, we briefly summarize the deterministic UC and SUC modeling approaches and then discuss their limitations. 
	Traditionally, scheduling and dispatch in power system operations have been done by using deterministic methods, and this is still the industry practice in most regions\cite{zhou2016stochastic}. Some recent studies also use the deterministic UC to analyze the impact of renewable resources on power system operations  \cite{ummels2007impacts,delarue2008adaptive}. 
	
	Basically, given the historical data, various methods can be used for wind power forecasting, such as persistence-based forecasting method \cite{Nielsen_etal_1998, Gneiting_etal_2007, Kavasseri_2009} and ARMA model\cite{ummels2007impacts}. The \textit{deterministic UC} makes the optimal decision, denoted by $\mathbf{u}^{\star, d}$, based on the point predictor of future wind power generation in the planning horizon, denoted by $\pmb{\hat{\xi}}$, by solving the deterministic optimization,
		\begin{eqnarray}
	\small
	\min_{\mathbf{u}} G^p(\mathbf{u}) & \equiv & \mathbf{C}^{su} \mathbf{u} +  \min_{\mathbf{y}} \mathbf{C}^{fuel}  \mathbf{y}(\mathbf{u},\pmb{\hat{\xi}})
	\label{eq: objdeter} \\
	\mbox{s.t. }  && A\mathbf{u} \leq B  
	\nonumber \\
	&& H\mathbf{u}+Q\mathbf{y}(\mathbf{u},\pmb{\hat{\xi}}) \leq M(\pmb{\hat{\xi}}).
	\nonumber 
	\end{eqnarray} 
	%Due to the complex constraints, the situation that a feasible solution cannot be obtained may occur. 
	%The author in \cite{guan2005conditions} proposed the analytical and computational necessary and sufficient conditions to determine the feasible unit commitment solution with constraints. In \cite{ozturk2004solution}, a heuristic is used to derive a feasible solution by adjusting the commitment state (e.g. turning on the cheapest generator available).
\textit{Thus, deterministic UC does not consider the prediction risk induced by both wind power generation inherent stochastic uncertainty and input model estimation uncertainty}. It will deliver an inferior  and unreliable decision, especially when the penetration of renewable energy increases. % This normally leads to a conservative operation with high operating costs or to an unanticipated high risk during operation \cite{bessa2014handling}. 
	
	%%%%%%%%%empirical UC
	%In contrast, the empirical SUC determines the UC decisions to provide potential reserves by considering a set of forecast scenarios. 
	%Thus, the SUC formulation includes multiple wind power scenarios and minimizes the expected cost across all scenarios. 
	%Different scenarios are assigned with different probabilities. Hence, the produced schedule can hedge against wind variations within the properly constructed scenario set. 
    Differing with the deterministic UC, the \textit{empirical SUC} considers the impacts from stochastic uncertainty of wind power generation on electric power systems \cite{Tuohy_etal_2009, Wang_Guan_Wang_2012,Huang_Zheng_Wang_2014}. Basically, given the historical data $\mathcal{D}$, to find the optimal decision, the empirical SUC takes the input model estimate, denoted by  $F^e$, of unknown $F^c$ as the true one, and then solves the stochastic optimization,
 	\begin{eqnarray}
	\small
	\min_{\mathbf{u}} G^p(\mathbf{u}) & \equiv & \mathbf{C}^{su} \mathbf{u} + \mbox{E}_{\pmb{\xi} \sim F^e} \left[ \min_{\mathbf{y}} \mathbf{C}^{fuel} \mathbf{y}(\mathbf{u},\pmb{\xi})\right]  
	\label{eq: objemp} 
 \\
	\mbox{s.t. }  && A\mathbf{u} \leq B   
	\nonumber \\
	&& H\mathbf{u}+Q\mathbf{y}(\mathbf{u},\pmb{\xi}) \leq M(\pmb{\xi}).	\nonumber
	\end{eqnarray} 
    Then, a Monte Carlo approach can be used to generate a finite number of scenarios from $F^e$, use the sample mean to approximate the expected future cost in the planning horizon, and obtain the optimal decision, denoted by $\mathbf{u}^{\star, e}$.
	\textit{Thus, the empirical UC ignores the input model estimation uncertainty and finite sampling error from SAA.} It could lead to an inferior and unreliable unit commitment decision, especially under the situations when the wind power penetration is high and the amount of representative real-world wind power historical data is limited; see as shown in the case studies in Section~\ref{sec:emp}.

	% =================== 
	\subsection{Data-Driven Stochastic Unit Commitment Model}
	\label{subsec:compounduc}

	\textit{In this paper, we propose a data-driven SUC accounting for both stochastic uncertainty of wind power generation and input model estimation uncertainty}. Given the historical data $\mathcal{D}$, the posterior distribution of underlying input model $F^c$ is used to quantify the model estimation uncertainty. Then, {the posterior predictive distribution, denoted by $F^p$, can quantify the prediction risk induced from both sources of uncertainties.} Thus, the proposed data-driven SUC model becomes,
	\begin{eqnarray}
	\small
	\min_{\mathbf{u}} G^p(\mathbf{u}) & \equiv & \mathbf{C}^{su} \mathbf{u} + \mbox{E}_{\pmb{\xi} \sim F^p} \left[ \min_{\mathbf{y}} \mathbf{C}^{fuel} \mathbf{y}(\mathbf{u},\pmb{\xi})\right]  
	\label{eq: objcom} \\
	\mbox{s.t. } &&  A\mathbf{u} \leq B  
	\nonumber  \\
	&& H\mathbf{u}+Q\mathbf{y}(\mathbf{u},\pmb{\xi}) \leq M(\pmb{\xi}).
	\nonumber 
	\end{eqnarray} 
	It can lead to cost-efficient, reliable and robust optimal unit commitment decision, denoted by $\mathbf{u}^{\star, p}$, which can hedge against the prediction risk induced by stochastic uncertainty of wind power generation and input model estimation uncertainty. %Our approach can be applied to the situations either we know the distribution family of $F^c$ or not.

	Given the historical data $\mathcal{D}$, the posterior distribution characterizing the input model estimation uncertainty can be obtained by the Bayes' rule, 
	$
	p(F|\mathcal{D})\propto p(F)p(\mathcal{D}|F),
	$
	where $p(F)$ denotes the prior and $p(\mathcal{D}|F)$ denotes the likelihood of historical data.
	Then, the density of posterior predictive distribution $F^p$,
	\begin{equation}
	f^p(\pmb{\xi}) \equiv \int p(\pmb{\xi}  |F) p(F|\mathcal{D}) d F,
	\label{eq.postPred}
	\end{equation}
	can quantify the overall prediction uncertainty of wind power generation with $p(F|\mathcal{D})$ characterizing the input model estimation uncertainty and $p(\pmb{\xi}  |F)$ characterizing the prediction uncertainty induced by wind power generation inherent volatility or stochastic uncertainty. 
	
	The proposed data-driven SUC can be applied to situations where the parametric family of underlying input model $F^c$ is known, e.g., \cite{Wang_Guan_Wang_2012, Zhang_Wang_Wang_2014, Pandzic_etal_2016}.
	In Sections~\ref{subsubsec: sixpacompound} and \ref{subsec:118bus_parametric}, when we study the six- and 118-bus power grid systems, we use the normal distribution assumption for illustration. 
	The proposed data-driven SUC in (\ref{eq: objcom}) can also apply to the situations where there is no strong prior information on the distribution family for $F^c$. In Sections~\ref{subsubsec: sixpaIMSAR} and \ref{subsec:118bus_nonparametric}, we suppose that there is no strong parametric assumption on the underlying input model $F^c$ and use the Bayesian nonparametric probabilistic forecast introduced in our previous study \cite{Xie_etal_2018} as an illustration.

	\section{OPTIMIZATION AND SELECTION}
	\label{sec:framework}
	
	%For complex power grids, it can be computationally expensive to search for optimal unit commitment decision. 
	When we solve the stochastic unit commitment optimization, a finite number of scenarios is typically used to approximate the expected cost in the planning horizon. It induces the \textit{finite sampling error}, which can be large especially for complex power grids with high wind power penetration. 
	In this section, we develop \textit{an optimization and selection (OPSEL) approach} which can utilize the parallel computing to quickly solve the data-driven SUC and find the optimal decision accounting for the impact of finite sampling error induced by SAA. This study can be applied to general stochastic UC problems and it can accelerate the real-time reliable decision making. 
	
	Since the time used to assess the performance of any given unit commitment decision is much less %that 
	than the optimization search, \textit{we propose an optimization solution methodology including two parts: optimization search and best candidate decision selection.} For the optimization search part as presented in Section~\ref{subsec: SAA}, we simultaneously solve $L$ independent SAA approximated data-driven SUC problems through parallel computing. It returns a set of optimal candidate solutions quantifying the impact of finite sampling error induced by SAA. For the best candidate decision selection part as shown in Section~\ref{subsec: OCBA}, we apply the ranking-and-selection methodology to efficiently allocate more computational budget to the most promising candidate decisions, improve the assessment of their performance, and quickly select the best decision. \textit{Thus, the combination of data-driven SUC and OPSEL can lead to the optimal and reliable unit commitment decision, which can hedge against: (1) inherent stochastic uncertainty of wind power generation, (2) input model estimation uncertainty, and (3) finite sampling error induced by SAA.}

	\subsection{Parallel Optimization Search Accounting for SAA Finite Sampling Error}
	\label{subsec: SAA}
	
	The second-stage expected economic dispatch cost $\mbox{E}_{\pmb{\xi} \sim F^p} \left[ \min_{\mathbf{y}} \mathbf{C}^{fuel} \mathbf{y}(\mathbf{u},\pmb{\xi})\right]  $
	does not have a closed-form expression, and it is typically approximated by using sample average approximation (SAA) (see the introduction of SAA in \cite{shapiro_2009}), which introduces the finite sample error. Specifically, we generate $S$ scenarios, $\pmb{\xi}^s\stackrel{i.i.d.}\sim F^p$ for $s=1,2,\ldots,S$, and use them to estimate the expected cost in the planning horizon. The \textit{SAA approximated data-driven SUC} in (\ref{eq: objcom}) becomes
	\begin{eqnarray}
	\small
	\min_{\mathbf{u}} & &\bar{G}^p(\mathbf{u})  =   \mathbf{C}^{su} \mathbf{u} +  \frac{1}{S}\sum_{s=1}^{S } \left[ \min_{\mathbf{y}} \mathbf{C}^{fuel} \mathbf{y}(\mathbf{u},\pmb{\xi}^s)\right]  
	\label{eq: genapp} \\
	\mbox{s.t. } && A\mathbf{u} \leq B 
	\nonumber  \\
	&& H\mathbf{u}+Q\mathbf{y}(\mathbf{u},\pmb{\xi}^s) \leq M(\pmb{\xi}^s)  \quad \mbox{for } s=1,2,\ldots, S.
	\nonumber 
	\end{eqnarray} 
	To accurately approximate the expected second-stage cost, the number of scenarios $S$ needs to be sufficiently large. It can be computationally infeasible to solve for complex power grids with high wind power penetration, especially when the quick decision making is required for intra-day unit commitment problems. Thus, SAA in (\ref{eq: genapp}) typically introduces the \textit{unignorable finite sampling error}.

	To efficiently employ the computational resource and \textit{quickly} deliver the optimal UC decision hedging against various sources of uncertainty, we exploit the parallel computing with $L$ available CPUs. With each $\ell$-th CPU, we first generate an independent set consisting of $S$ scenarios, $\pmb{\xi}^1,\ldots,\pmb{\xi}^S$, drawn from $F^p$, and then we solve the corresponding  SAA approximated data-driven SUC problems in (\ref{eq: genapp}). % through parallel computing.  Specifically, the $\ell$-th CPU is used to solve the SAA approximated data-driven SUC problem with the $\ell$-th set of scenarios. 
	In the case study, we use the L-Shaped algorithm for optimization \cite{Zheng_etal_2013, Zheng_Wang_Liu_2015}. %Yet, our approach can be applied to other algorithms as well, such as the progressive hedging (PH) algorithm \cite{Ryan_etal_2013, Cheung_etal_2015} and the linear sensitivity factor (LSF) method \cite{Wu_2013, Wang_etal_2016}. 
	\textit{Thus, based on the parallel optimization search, we obtain the optimal unit commitment candidate decisions, denoted by  $\widehat{\mathbf{u}}^{\star }_{\ell}$ with $\ell=1,\ldots,L$, quantifying the impact of finite sampling error.}

	\subsection{Ranking and Selection based Best Decision Selection}
	\label{subsec: OCBA}

	%%%%%%%%OCBA

	\begin{sloppypar}
	To eliminate the impact of finite sampling error quantified by candidates $\widehat{\mathbf{u}}^{\star}_{1}, \ldots, \widehat{\mathbf{u}}^{\star}_{L}$, we efficiently utilize the computational resource to assess the performance of candidate decisions, $ G^p(\widehat{\mathbf{u}}^{\star }_{1}), \ldots, G^p(\widehat{\mathbf{u}}^{\star}_{L})$, and select the best one, 
	\[
	{\mathbf{u}}^{\star}_b \equiv \mbox{argmin}_{\widehat{\mathbf{u}}^{\star}_{\ell} \in \{\widehat{\mathbf{u}}^{\star}_{1}, \ldots, \widehat{\mathbf{u}}^{\star}_{L}\}} G^p(\widehat{\mathbf{u}}^{\star }_{\ell}).
	\] 
	with the subscript $b$ denoting the best unit commitment decision.
	The number of CPUs, $L$, could be large. 
	For each candidate, the performance $G^p(\widehat{\mathbf{u}}^{\star}_{\ell})$ with $\ell=1,2,\ldots,L$ can be assessed by using Monte Carlo approach.
	\textit{We sequentially allocate the computational budget to the most promising candidates $\widehat{\mathbf{u}}^{\star }_{\ell}$ so that we can provide more accurate estimation of their expected cost $G^p(\widehat{\mathbf{u}}^{\star }_{\ell})$ and efficiently select out the best solution.} 
	\end{sloppypar}

	Basically, built on the results from optimal search described in Section~\ref{subsec: SAA}, we further use the OCBA approach \cite{Chen_etal_2000} to guide the sequential allocation of computational resource to promising candidates. % and correctly assess their performance. 
	Here, the each unit of computational budget is measured by the cost of solving one second-stage economic dispatch problem for each scenario.
	Let $N_{k, \ell}$ be the \textit{accumulated} number of scenarios assigned to the candidate solution $\widehat{\mathbf{u}}^{\star}_{\ell}$ for $\ell =1, \ldots, L$
	until the $k$-th iteration of sequential candidate selection, and the objective estimate is 
		\begin{equation}
	\bar{G}^p_k(\widehat{\mathbf{u}}_\ell^\star)  =   \mathbf{C}^{su} \widehat{\mathbf{u}}_\ell^\star +  \frac{1}{N_{k, \ell}}\sum_{s=1}^{N_{k, \ell}} \left[ \min_{\mathbf{y}} \mathbf{C}^{fuel} \mathbf{y}(\widehat{\mathbf{u}}_\ell^\star,\pmb{\xi}^s)\right].
	\label{eq.G_bar}
	\end{equation}
	With more scenarios, we can estimate the performance ${G}^p_k(\widehat{\mathbf{u}}_\ell^\star)$ more accurate.
	The number of initial scenarios is $N_{0, \ell}=S$ since the data-driven SUC approximated with $S$ samples is solved by the $\ell$-th CPU to obtain $\widehat{\mathbf{u}}^{\star}_{\ell}$; see Eq.~(\ref{eq: genapp}). 
	At the $k$-th iteration, we allocate $\Delta T$ additional scenarios to the candidates $\widehat{\mathbf{u}}^{\star}_{\ell}$ with $\ell =1, \ldots, L$. Then, we solve the corresponding economic dispatch problems for the new generated scenarios and update the objective estimate for $ G^p(\widehat{\mathbf{u}}^{\star }_{\ell})$ by using Eq.~(\ref{eq.G_bar}).
	
	Specifically, based on  \cite{Chen_etal_2000}, the optimal budget allocation $N_{k, \ell}$ is obtained by solving
	\begin{eqnarray}
	\frac{N_{k,\ell}}{N_{k, \ell^\prime}} &= &\left( \frac{\delta_{k-1,\ell^\prime}}{\delta_{k-1,\ell}} \right)^2 \mbox{ for } \ell, \ell^\prime \neq b   
	\label{eq: ocba1}  \label{eq.ratio_N}  \\
	N_{k,b} &=&\widehat{\sigma}_{k-1,b} \sqrt{\sum_{\ell=1, \ell\neq b}^{L} \frac{N^2_{k,\ell}}{\widehat{\sigma}^2_{k-1,\ell}}} 
	\label{eq: ocba2} \nonumber \\
	\sum_{\ell =1}^{L} N_{k,\ell} &=&  L \times S +  k \times \Delta T 
	\nonumber 
	\label{eq: ocba3}
	\end{eqnarray}    
	where $N_{k,b}$ denotes the number of scenarios assigned to the current best candidate selected from the $(k-1)$-th iteration
	\begin{equation}
	\widehat{\mathbf{u}}^{\star}_{k-1,b} \equiv \mbox{argmin}_{\widehat{\mathbf{u}}^{\star}_{\ell} \in \{\widehat{\mathbf{u}}^{\star}_{1}, \ldots, \widehat{\mathbf{u}}^{\star}_{L}\}} \bar{G}^p_{k-1}(\widehat{\mathbf{u}}^{\star }_{\ell}).
	\label{eq.bestEst}
	\end{equation}
	\begin{comment}
	and the objective estimate is 
	\begin{equation}
	\bar{G}_{k-1}^p(\widehat{\mathbf{u}}^{\star}_{\ell}) =  \mathbf{C}^{su}\widehat{\mathbf{u}}^{\star}_{\ell}+  \frac{1}{N_{k-1,\ell}}\sum_{s=1}^{N_{k-1, \ell} } \left[ \min_{\mathbf{y}} \mathbf{C}^{fuel} \mathbf{y}(\widehat{\mathbf{u}}^{\star}_{\ell}, \pmb{\xi}^s)\right].
	\label{eq: updatemean}
	\end{equation} 
	\end{comment}
	The estimate of variance, ${\sigma}_\ell^2=\mbox{Var}\left[ \min_{\mathbf{y}} \mathbf{C}^{fuel} \mathbf{y}(\mathbf{u},\pmb{\xi}) \right]$, is the sample variance obtained from the $(k-1)$-th iteration,
	\begin{eqnarray}
	\widehat{\sigma}_{k-1,\ell}^2 = \frac{1}{N_{k-1,\ell}-1} \sum_{s=1}^{N_{k-1,\ell}} \Big[     \mathbf{C}^{su}\widehat{\mathbf{u}}^{\star}_{\ell} \Big.
\Big. +    \min_{\mathbf{y}} \mathbf{C}^{fuel} \mathbf{y}(\widehat{\mathbf{u}}^{\star}_{\ell},\pmb{\xi}^s)   
	- \bar{G}^p_{k-1}  (\widehat{\mathbf{u}}^{\star }_{\ell})  \Big]^2
	\label{eq.sigmaHat} \nonumber 
	\end{eqnarray}
	and   
	$\delta_{k-1,\ell}$ denotes the standardized distance between $\widehat{\mathbf{u}}^{\star}_{\ell}$ with the current estimated best candidate $\widehat{\mathbf{u}}^{\star}_{k-1,b}$,
	\begin{equation}
	\delta_{k-1,\ell} \equiv \frac{ \bar{G}^p_{k-1}(\widehat{\mathbf{u}}^{\star }_{\ell})-\bar{G}^p_{k-1}(\widehat{\mathbf{u}}^{\star }_{k-1,b})}{\widehat{\sigma}_{k-1,\ell}}.
	\label{eq: sdis} 
	\end{equation} 
	\textit{By applying Eq.~(\ref{eq.ratio_N}) and (\ref{eq: sdis}), the budget allocation to any candidate $\widehat{\mathbf{u}}^{\star}_{\ell}$ is directly related to its standardized distance with the current best estimate, which can allocate more computational resource to the promising candidates and efficiently select out the best decision.}
	Thus, in the $k$-th iteration, the number of new scenarios allocated to  the candidate $\widehat{\mathbf{u}}^{\star}_{\ell}$ for $\ell=1,2,\ldots, L$ is,
	\begin{equation}
	\Delta N_{k, \ell} = \max (0,  N_{k, \ell}- N_{k-1, \ell}).
	\label{eq: newsample}
	\end{equation} 
	Then, we solve the additional $\Delta N_{k, \ell}$ second-stage economic dispatch problems and update the objective estimate $
	\bar{G}_k^p(\widehat{\mathbf{u}}^{\star}_{\ell})$.
	
	The OPSEL procedure is provided in Algorithm~\ref{al: OCBA}. %We first specify $S$ for approximated data-driven SUC in (\ref{eq: genapp}) and 
	We denote the overall computational budget in terms of number of scenarios allocated for the candidate selection by $T$. In Step~(1), we utilize $L$ CPUs to simultaneously solve the sample average approximated SUC problems with form (\ref{eq: genapp}) and obtain the optimal candidate decisions $\widehat{\mathbf{u}}^{\star}_{\ell}$ with $\ell=1,2,\ldots,L$. Then, in Steps~(2) and (3), we sequentially allocate the computational resource to $\widehat{\mathbf{u}}^{\star}_{1}, \ldots, \widehat{\mathbf{u}}^{\star}_{L}$ and select the best candidate decision hedging against the finite sampling error.

	\begin{algorithm}
		
		Step~(0) Specify $S$ and the total budget $T$, the total number of scenarios used for the candidate solution selection.
		
		Step~(1) Utilize $L$ CPUs to simultaneously solve the SUC problems (\ref{eq: genapp}) approximated by $S$ scenarios, and obtain the optimal candidate decisions $\widehat{\mathbf{u}}^{\star}_{1}, \ldots, \widehat{\mathbf{u}}^{\star}_{L}$.  
		
		Step~(2) At the $k$-th iteration of selection, allocate $\Delta T$ new scenarios to $\widehat{\mathbf{u}}^{\star}_{1}, \ldots, \widehat{\mathbf{u}}^{\star}_{L}$ by using~(\ref{eq: newsample}) and solve the additional second-stage dispatch problems. Update $\bar{G}^p_k(\widehat{\mathbf{u}}^{\star }_{\ell})$ for $\ell=1, \ldots, L$ and $\widehat{\mathbf{u}}^{\star}_{k,b}$ by applying (\ref{eq.G_bar}) and (\ref{eq.bestEst}).

		Step~(3) Repeat Step~(2) until reaching to the budget $T$. Return $\widehat{\mathbf{u}}^{\star}_{k,b}= \mbox{argmin}_{\widehat{\mathbf{u}}^{\star}_{\ell} \in \{\widehat{\mathbf{u}}^{\star}_{1}, \ldots, \widehat{\mathbf{u}}^{\star}_{L}\}} \bar{G}^p_{k}(\widehat{\mathbf{u}}^{\star }_{\ell}).$

		\caption{The Optimization and Selection Procedure}
		\label{al: OCBA}
	\end{algorithm}

\vspace{-0.4in}

\section{Empirical Studies}
\label{sec:emp}

We use the six-bus system from \cite{Wang_etal_2017} in Section~\ref{subsec:sixBus} and the derivative 118-bus system from \cite{pena2017extended} in Section~\ref{subsec:118Bus} to compare the performance of proposed data-driven SUC framework with the deterministic unit commitment (UC) and the empirical SUC %uner 
under the situations when the parametric input model for wind power generation is known or not. 
In the empirical studies, we consider the two-stage SUC problem,
\begin{equation}
    \begin{aligned}
        \min & \quad  G(u_{i,t}) =\sum_{t=1}^{T} \sum_{i=1}^{I}  (C^i F_{mi} u_{i,t} + SU_{i,t} +SD_{i,t} )  \\
        & + \mbox{E} \left[ \min_{P_{i,t}, P_{b,t}^{ens}, P_{w,t}^{wc} } \sum_{t=1}^{T} \sum_{i=1}^{I}C^i F_{ai} P_{i,t} +\sum_{t=1}^{T} \sum_{b=1}^{B} C^{ens} P_{b,t}^{ens}
        +\sum_{t=1}^{T} \sum_{w=1}^{W} C^{wc}  P_{w,t}^{wc}\right]
   \end{aligned}
   \label{eq: sixbusobj}
\end{equation}   

\begin{comment}
\begin{eqnarray}
    \label{eq: sixcons2}
    \textrm{s.t.} \quad \quad \quad \quad \quad \quad \quad \quad \quad     u_{i,t} -u_{i,t-1}  \leq u_{i,k}  \quad \forall k=t, \ldots, \min(T, t+T_{i}^{on}-1)  \\
  	\label{eq: sixcons3}
    u_{i,k}  \leq 1 +u_{i,t} -u_{i,t-1}  \quad \forall k=t, \ldots, \min(T, t+T_{i}^{off}-1)  \\
   \label{eq: sixcons4} 
        \sum_{i=1}^{I} P_{i,t} +\sum_{w=1}^{W} (P_{w,t}^c-P_{w,t}^{wc})=\sum_{b=1}^{B} (P^D_{b,t} - P_{b,t}^{ens} )  \\
      \label{eq: sixcons5}
         -PL_{\ell,\max} \leq \sum_{b=1}^{B} G_{\ell-b} P^D_{b,t} +\sum_{i=1}^{I} G_{\ell-i} P_{i,t} + \sum_{w=1}^{W} G_{\ell-w}(P_{w,t}^c-P_{w,t}^{wc}) \leq PL_{\ell,\max}     \\  
   \label{eq: sixcons6} 
        u_{i,t}P_{i,\min} \leq u_{i,t}P_{i,t} \leq u_{i,t}P_{i, \max}  \quad \forall i, \quad \forall t   \\
%   \label{eq: sixcons7}
%        u_{i,k} P_{i,k} = u_{i,k} \cdot \left[ u_{i,t-1} P_{i,t-1} +(1-u_{i,t-1})  P_{i,\min} \right] \quad \forall k=t, \ldots, \min(T, t^\prime) \\
        \quad \quad u_{i,t}  
        \quad \mbox{binary}   
        \label{eq: sixcons1}
\end{eqnarray}
\end{comment}

\begin{align}
    \textrm{s.t.} \quad \quad \quad \quad \quad \quad  \quad  \quad  \quad    u_{i,t} -u_{i,t-1}  \leq u_{i,k}  \quad \forall k=t, \ldots, \min(T, t+T_{i}^{on}-1)      \label{eq: sixcons2} \\
    u_{i,k}  \leq 1 +u_{i,t} -u_{i,t-1}  \quad \forall k=t, \ldots, \min(T, t+T_{i}^{off}-1)    	\label{eq: sixcons3}\\
    \sum_{i=1}^{I} P_{i,t} +\sum_{w=1}^{W} (P_{w,t}^c-P_{w,t}^{wc})=\sum_{b=1}^{B} (P^D_{b,t} - P_{b,t}^{ens} )     \label{eq: sixcons4}  \\
    -PL_{\ell,\max} \leq \sum_{b=1}^{B} G_{\ell-b} P^D_{b,t} +\sum_{i=1}^{I} G_{\ell-i} P_{i,t} + \sum_{w=1}^{W} G_{\ell-w}(P_{w,t}^c-P_{w,t}^{wc}) \leq PL_{\ell,\max}          \label{eq: sixcons5} \\  
    u_{i,t}P_{i,\min} \leq u_{i,t}P_{i,t} \leq u_{i,t}P_{i, \max}  \quad \forall i, \quad \forall t      \label{eq: sixcons6}  \\
%   \label{eq: sixcons7}
%        u_{i,k} P_{i,k} = u_{i,k} \cdot \left[ u_{i,t-1} P_{i,t-1} +(1-u_{i,t-1})  P_{i,\min} \right] \quad \forall k=t, \ldots, \min(T, t^\prime) \\
        \quad \quad u_{i,t}  
        \quad \mbox{binary}   
        \label{eq: sixcons1}
\end{align}

The objective in (\ref{eq: sixbusobj}) is to minimize the total expected cost occurring in the planning horizon $T$, including the start-up cost $SU_{i,t}$, the turn-off cost $SD_{i,t}$ and minimal thermal operation cost $C^i F_{mi} u_{i,t}$ incurring in the fist stage, where $C^i$ is the fuel price and $F_{mi}$ is the amount of fuel consumption for the minimal output for generator $i$. %, incur at the fist stage and the remaining three incur at the second stage. % The start-up cost $SU_{i,t}$ incurs when the generator $i$ is turned on, and it is defined as $SU_{i,t} \equiv P^{on}_{i} \max(u_{i,t}-u_{i,t-1}, 0)$, where $P^{on}_{i}$ is the start-up cost per time for generator $i$.
%If generator $i$ is not committed at hour $(t-1)$, i.e., $u_{i,t-1}=0$ while it is committed at hour $t$, i.e., $u_{i,t}=1$. Then, $u_{i,t}-u_{i,t-1}=1$. Hence, a start-up cost for generator $i$ happens at hour $t$. Similarly,
%The turn-off cost $SD_{i,t}$ happens when generator $i$ is shut-down and $SD_{i,t} \equiv P^{off}_{i} \max(u_{i,t-1}-u_{i,t}, 0)$, where $P^{off}_{i}$ is the turn-off cost per time for generator $i$. 
%If a generator is committed at hour $(t-1)$, i.e., $u_{i,t-1}=1$ while it is not committed at hour $t$, i.e., $u_{i,t}=0$. Then, the generator is shut down at hour $t$ and a turn-off cost incurs. 
%In addition, once thermal generator $i$ is committed, it has to produce above the minimal production level. Thus, a minimal thermal operation cost incurring at the first-stage is denoted by $C^i F_{mi} u_{i,t}$, where $C^i$ is the fuel price and $F_{mi}$ is the amount of fuel consumption for the minimal output for generator $i$. %\shortcite{Wang_etal_2017}. 
%The remaining three costs incur at the second-stage. 
Since the thermal generator consumes the extra fuel to produce $P_{i,t}$, the additional production cost in the second-stage is $C^i F_{ai} P_{i,t}$, where $F_{ai}$ is the fuel consumption 
needed to generate $P_{i,t}$ power production. 
%If the smart grid could not produce enough energy to satisfy the load demand, a shortage penalty may incur. 
{The penalty cost of non-satisfied demand} for bus $b$ at time period $t$ is $C^{ens}P_{b,t}^{ens}$, where $C^{ens}$ is the unit load shedding price and $P_{b,t}^{ens}$ is the amount of unmet load at bus $b$ in time period $t$.  Lastly, for the wind  power,  we may not use up all its capacity $P_{w,t}^{c}$ and there exists a wind farm curtailment $P_{w,t}^{wc}$. %Since the monetary incentives are provided for wind power production, we loss the monetary reward if we curtail wind farm production. % \shortcite{Wang_etal_2017}
The wind curtailment cost at the $t$-th hour for wind farm $w$ is $C^{wc}  P_{w,t}^{wc}$, where $C^{wc}$ is the per unit monetary reward for the wind production and $P_{w,t}^{wc}$ is the amount of wind curtailment.
 Constraints (\ref{eq: sixcons2})--(\ref{eq: sixcons3}) formulate the minimum up and down time requirement, Constraints (\ref{eq: sixcons4})--(\ref{eq: sixcons5}) are for the nodal power balance and the DC power flow constraint, and Constraint (\ref{eq: sixcons6}) enforces the minimum and maximum generator output limits. 

\subsection{Empirical Study with Six-Bus System}
\label{subsec:sixBus}

\begin{figure}[h!]
	\vspace{-0.15in}
	{
		\centering
		\includegraphics[scale=0.6]{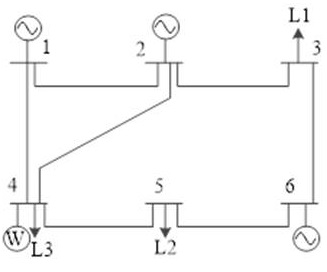}
		\vspace{-0.2in}
		\caption{Six-Bus System}
		\label{fig:Figure1}
	}
\end{figure}

%In this subsection, we use the six-bus system in \cite{Wang_etal_2017} to study the performance of our approach. 
For the six-bus system, it consists of three thermal units, a wind farm and seven transmission lines as depicted in Fig.~\ref{fig:Figure1}. The three thermal units are located in No.1, No.2 and No.6 buses, while No.4 Bus hosts a wind farm; see Table~\ref{table: sixbus} for the description of the six-bus system.  Tables~\ref{table: generator} and \ref{table: fuelconsumption} describe the characteristics of the thermal generators, while the characteristics of the transmission lines are provided in Table~\ref{table: transmissionline}.  
Since the uncertainty in wind supply typically dominates, in this study, we assume deterministic loads and stochastic wind supply \cite{Zhou_etal_2013}. Following \cite{Wang_etal_2017}, we use the 2006 data of the U.S. Illinois power system for the load $P_{b,t}^D$ and the wind supply $P_{w,t}^c$. Then, the \textit{wind penetration level} measured by the ratio of wind power generation to actual demand, $R = \sum_{w=1}^W\sum_{t=1}^T{P_{w,t}^c} /\sum_{b=1}^B\sum_{t=1}^T{P_{b,t}^D}$, is 37.8\%.  
%For the specific two-stage SUC problem, we follow the formulation in \cite{Wang_etal_2017}. %In this model, we consider five different types of costs, including the start-up cost, the shutdown cost, the thermal generator production cost, the wind curtailment penalty cost and the load shedding penalty cost; see  \cite{Wang_etal_2017} for the detail. 
%The first three types of costs are related to thermal generator and the corresponding fuel consumption costs are given in Table~\ref{table: fuelconsumption}. 
Additionally,  the wind curtailment price is fixed at $50\$/MWh$ and the load shedding price is set at $3500\$/MWh$ \cite{Wang_etal_2017}.
%Six constraints are included in the unit commitment model, which are expressed in Equations~(4)-(9) in \cite{Wang_etal_2017}. The ramp-up and ramp-down constraint expressed in Equation~(10) in \cite{Wang_etal_2017} is relaxed for the ease of our discussion. 

\vspace{-0.0in}

\begin{table}[]
	\centering
	\caption{Bus Data}
	\label{table: sixbus}
	\begin{tabular}{|l|l|l|l|l|}
		\hline
		Bus ID & Type    & Thermal Unit & Wind Farm & Load Share \\ \hline
		NO 1     & Thermal & G1           &           &            \\ \hline
		NO 2     & Thermal & G2           &           &            \\ \hline
		NO 3     &         &              &           & 20\%       \\ \hline
		NO 4     & Wind    &              & W1        & 40\%       \\ \hline
		NO 5     &         &              &           & 40\%       \\ \hline
		NO 6     & Thermal & G3           &           &            \\ \hline
	\end{tabular}
\end{table}

\begin{table}[]
	\centering
	\caption{Thermal Generator Data}
	\label{table: generator}
	\scalebox{0.9}{
		\begin{tabular}{|l|l|l|l|l|l|}
			\hline
			Unit & Pmax(MW) & Pmin(MW) & Ini.State (h) & Min Off(h) & Min On (h) \\ \hline
			G1 & 220 & 90 & 4 & -4 & 4 \\ \hline
			G2 & 100 & 20 & 2 & -3 & 2 \\ \hline
			G3 & 30 & 10 & -1 & -1 & 1 \\ \hline
	\end{tabular} }
\end{table}

% \usepackage{multirow}
% \usepackage{multirow}
\begin{table}[]
	\centering
	\caption{Thermal Generator Data}
	\label{table: fuelconsumption}
	% \usepackage{multirow}
	\scalebox{0.83}{
		\begin{tabular}{|l|l|l|l|l|l|l|}
			\hline
			\multirow{2}{*}{Unit} & \multicolumn{3}{l|}{Fuel consumption Function}                                                                                                                          & \multirow{2}{*}{\begin{tabular}[c]{@{}l@{}}Start up\\  Fuel \\ (MBtu)\end{tabular}} & \multirow{2}{*}{\begin{tabular}[c]{@{}l@{}}Shut down\\  Fuel\\ (MBtu)\end{tabular}} & \multirow{2}{*}{\begin{tabular}[c]{@{}l@{}}Fuel\\ Price \\ (\$)\end{tabular}} \\ \cline{2-4}
			& \begin{tabular}[c]{@{}l@{}}a \\ (MBtu)\end{tabular} & \begin{tabular}[c]{@{}l@{}}b \\ (MBtu/MWh)\end{tabular} & \begin{tabular}[c]{@{}l@{}}c\\ (MBtu/MW2h)\end{tabular} &                                                                                     &                                                                                     &                                                                               \\ \hline
			G1        & 176.9       & 13.5         & 0.0004         & 180         & 50     & 1.2469    \\ \hline
			G2        & 129.9      & 32.6         & 0.001           & 360         & 40     & 1.2461     \\ \hline
			G3       & 137.4   & 17.6           & 0.005         & 60              & 0  & 1.2462         \\ \hline
	\end{tabular} }
\end{table}

\begin{table}[]
	\centering
	\caption{Transmission Line Data}
	\label{table: transmissionline}
	\scalebox{0.9}{
		\begin{tabular}{|l|l|l|l|l|}
			\hline
			Line No. & From Bus & To Bus & X (p.u) & Flow Limit (MW) \\ \hline
			1        & 1        & 2      & 0.17    & 200             \\ \hline
			2        & 1        & 4      & 0.258   & 100             \\ \hline
			3        & 2        & 4      & 0.197   & 100             \\ \hline
			4        & 5        & 6      & 0.14    & 100             \\ \hline
			5        & 2        & 3      & 0.037   & 200             \\ \hline
			6        & 4        & 5      & 0.037   & 200             \\ \hline
			7        & 3        & 6      & 0.018   & 200             \\ \hline
	\end{tabular} }
\end{table}

\subsubsection{Case Study with Parametric Forecast Model}
\label{subsubsec: sixpacompound}

In this section, suppose the input model of wind power generation $F^c$ follows the normal distribution \cite{Wang_Shahidehpour_Li_2008, Wang_Guan_Wang_2012, Pandzic_etal_2016}
with unknown parameters estimated by valid historical data. 
%We compare the performance of proposed data-driven SUC with existing deterministic UC and empirical SUC approaches.
Here, we consider the day-ahead unit commitment with the planning horizon equal to 24 hours.
In each day $d$, suppose that the wind power generation at the $t$-th hour follows a normal distribution, $\xi_{d,t} \sim N(\mu_{d,t}^c, \phi_{d,t}^2)$, where  $\mu_{d,t}^c$ and $\phi_{d,t}^2$ are mean and variance.
Thus, the underlying true input model $F^c$ for $\xi_{d,t}$ in SUC (\ref{eq: genobj}) is $N(\mu_{d,t}^c, \phi_{d,t}^2)$.
To evaluate the performance, we pretend that the true parameter $\mu_{d, t}^c$ is unknown. Suppose that wind power at the $t$-th hour in the past $m$ days follows the same distribution. To predict $\xi_{d,t}$, we use the valid historical observation
$\mathcal{D}_{d, t} \equiv \{ \xi^r_{(d-m), t}, \ldots, \xi^r_{(d-1),t}  \}$ with $t=1,\ldots,24$ hour, where $\xi^r_{(d-i),t}$ denotes the \textit{real} wind power observation at the $t$-th hour in day $(d-i)$ with $ i=1, \ldots, m$.
Here, we set $\mu_{d, t}^c$ equal to the 2006 wind power generation.

The empirical SUC takes the estimated input model coefficient, sample mean, as the true one. Thus, the input model $F^e_{d,t}$ for SUC is $N(\bar{\xi}^r_{d,t}, \phi_{d,t}^2)$, where the sample mean of the historical data 
$
\bar{\xi}^r_{d,t} = \frac{1}{m} \sum_{i=1}^{m} \xi^r_{(d-i),t}
$
is the plug-in estimate of unknown parameter $\mu^c_{d,t}$
For the proposed data-driven SUC, the model estimation uncertainty is characterized by the posterior distribution. Without strong information about the mean $\mu_{d, t}^c$, we use the non-informative prior, a normal distributed with mean zero and infinite variance \cite{Yang_Berger_1997}. The posterior distribution is $p(\mu_{d, t}|\mathcal{D}_{d, t}) = N(\bar{\xi}^r_{d,t}, {\phi_{d,t}^2}/{m})$. Then, the resulting posterior predictive distribution $F^p_{d,t}$ is $N(\bar{\xi}^r_{d,t}, (1+\frac{1}{m})\phi_{d,t}^2)$, which quantifies the prediction risk accounting for both input model estimation uncertainty and wind power generation inherent stochastic uncertainty,

We compare the performance of unit commitment decisions obtained from the data-driven SUC with the empirical SUC under various settings with standard deviation $\phi_{d,t} =5\%\mu_{d,t}^c, 10\%\mu_{d,t}^c, 20\% \mu_{d,t}^c$. Let $n_d$ denote the total number of days used for the evaluation. \textit{Let $\widehat{\mathbf{u}}_{d}^{\star, p}$ and $\widehat{\mathbf{u}}^{\star, e}_{d}$ denote the 24-hour optimal unit commitment decisions obtained by data-driven and empirical SUCs with $d=1, \ldots, n_d$.} Then, the total expected costs obtained by these approaches are 
\begin{equation}
\sum G^p =\sum_{d=1}^{n_d} G(\widehat{\mathbf{u}}_{d}^{\star, p})
\mbox{ and } 
\sum G^e =\sum_{d=1}^{n_d} G(\widehat{\mathbf{u}}_{d}^{\star, e}).
\label{eq.result_G}
\end{equation}

Since there is no closed form, the sample average approximations, $\bar{G}(\widehat{\mathbf{u}}^{\star, p}_{d})$ and $\bar{G}(\widehat{\mathbf{u}}^{\star, e}_{d})$, are used to estimate the true objectives. 
To determine a proper scenario size $S_e$ so that we can estimate $G(\widehat{\mathbf{u}}^{\star, p}_{d})$ and $G(\widehat{\mathbf{u}}^{\star, e}_{d})$ accurately, we conduct a side experiment and consider the high uncertainty case with $\phi_{d,t}=20\% \mu_{d,t}^c$. In addition, since the empirical approach ignores the input model parameter estimation uncertainty, % and directly estimates the parameters from the historical observations, 
the unit commitment decisions highly fluctuate with the random wind power observations and its solution quality is more volatile. Thus, to decide the proper sample size that can ensure the accurate estimation of the \textit{expected} total cost, we consider the empirical approach. Specifically, we first apply the L-shaped algorithm to solve the empirical SUC with the expected cost approximated by SAA having $S=50$ and obtain a unit commitment decision $\widehat{\mathbf{u}}_{d}^{\star, e }$. Then, we estimate the expected cost by using $S_e$ scenarios to obtain $\bar{G}(\widehat{\mathbf{u}}_{d}^{\star, e })$
and calculate the relative error, $\mbox{relativeError}\equiv {|\bar{G}(\widehat{\mathbf{u}}_{d}^{\star, e })-G(\widehat{\mathbf{u}}_{ d}^{\star, e })|}/{G(\widehat{\mathbf{u}}_{ d}^{\star, e })}$, where $G(\widehat{\mathbf{u}}_{d}^{\star, e })$ denotes the objective value obtained by using $10^5$ scenarios.  Suppose that $10^5$ is large enough so that the estimation error of $G(\widehat{\mathbf{u}}_{d}^{\star, e })$ is negligible. The maximum relative
error obtained from $10$ day-period  is recorded in Table~\ref{table:absolute}. We observe that $S_e=1000$ achieves accuracy with the maximum relative error not exceeding $1.0\%$. Balancing the computational cost and the accuracy, we use $S_e=1000$ to evaluate the true expected cost.

\vspace{-0.0in}

\begin{table}%
	\centering
	\caption{The Maximum Absolute Relative Difference for $G(\widehat{\mathbf{u}}_{d}^{\star e})$ Estimation
		\label{table:absolute}}{
		\scalebox{0.95}{
			\begin{tabular}{|c|c|c|c|c|c|c|} 
				\hline
				$S_e$  & $10^2$ & $5\times10^2$ & $10^3$  & $5\times10^3$ &$10^4$ &$5\times10^4$ \\  \hline
				relativeError       & 3.0\% & 1.5\% & 0.9\% & 0.9\% & 0.7\% & 0.1\%  \\ \cline{1-7}
	\end{tabular}} }
\end{table}%	

The wind power data in 2006 October are used for evaluation. Let $m =1$. In the study, we set the scenario size $S=50$ and get the optimal decision estimates $\widehat{\mathbf{u}}_{d}^{\star, e }$ and $\widehat{\mathbf{u}}_{d}^{\star, p }$ by solving the sample average approximated empirical and data-driven SUCs. %, i.e., only the previous day has the same hourly wind power distribution with today. % For each $d$ in the testing period, we record the true objectives of $\widehat{\mathbf{u}}^{\star }_{p,d}$ and $\widehat{\mathbf{u}}^{\star}_{ e,d}$, denoted by ${G}(\widehat{\mathbf{u}}^{\star}_{ p,d})$ and ${G}(\widehat{\mathbf{u}}^{\star }_{e,d})$ accordingly. 
For cases with $\phi_{d,t} =5\%\mu_{d,t}^c, 10\%\mu_{d,t}^c, 20\% \mu_{d,t}^c$, the results of $\sum G^p$ and $\sum G^e$ in (\ref{eq.result_G}) for the one-month period are recorded in Table~\ref{table: sixbusr}. 
We also record the relative expected saving obtained by our method, denoted by $r\Delta G$, 
	\[
	r\Delta G= \frac{\sum G^e-\sum G^p}{\sum G^p}.
	\]	

The results in Table~\ref{table: sixbusr} demonstrate that the proposed data-driven SUC significantly outperforms the empirical SUC. When $\phi_{d,t}= 5\% \mu_{d,t}^c$, the total \textit{expected} cost-saving by our approach is $208,730$, which represents a $8.8\%$ lower cost than the empirical SUC. As $\phi_{d,t}$ increases, the advantages of data-driven SUC tend to be larger. When $\phi_{d,t}= 20\% \mu_{d,t}^c$, our approach outperforms the empirical approach by $15.4\%$ savings. \textit{Thus, the proposed data-driven SUC can lead to better and more robust unit commitment decision and the advantage becomes larger when the wind penetration is higher.}

\begin{table}[]
	\centering
	\caption{Total costs obtained by data-driven SUC, empirical SUC, and deterministic UC approaches under the cases with different levels of wind power penetration}
	\label{table: sixbusr}
{
		\scalebox{1.1}{
			\begin{tabular}{|l|l|l|l|}
				\hline
				& $\sum G^p$ & $\sum G^e $ & $r\Delta G$  \\ \hline
				$\phi_{d,t}= 5\% \mu_{d,t}^c$   &      2,152,146        &    2,360,877  &  8.8\%       \\ \hline
				$\phi_{d,t}= 10\% \mu_{d,t}^c$   &   2,443,784          &  2,844,019     &  14.1\%     \\ \hline
				$\phi_{d,t}= 20\% \mu_{d,t}^c$   &     2,671,475          &    3,156,688  &  15.4\%     \\ \hline
		\end{tabular}}
	}
\end{table}

\vspace{-.05in}

\subsubsection{Case Study with Nonparametric Forecast Model}
\label{subsubsec: sixpaIMSAR}

In the real application, we often do not have any strong prior knowledge about the underlying input model $F^c$ and its distribution family is typically unknown. Thus, we consider a unit commitment problem with nonparametric forecast models.
Here, we compare the performance of various approaches, including (1)~the deterministic unit commitment (UC); (2)~the empirical SUC accounting for wind power stochastic uncertainty; and (3)~the data-driven SUC accounting for both wind power stochastic uncertainty and input model estimation uncertainty. 
Specifically, for the data-driven SUC, we use the Bayesian nonparametric wind power forecast model proposed in our previous study \cite{Xie_etal_2018}. For the empirical SUC and deterministic UC, we use probabilistic and deterministic persistence models \cite{Nielsen_etal_1998, Gneiting_etal_2007, Kavasseri_2009}.
At the time period $h_i$, the probabilistic persistence model is based on the empirical predictive distribution for $\pmb{\xi}_{h_t+i}$ specified by
\[
\{\pmb{\xi}^r_{h_t}-\pmb{\xi}^r_{h_t-j}+\pmb{\xi}^r_{h_t-j-i}: j=0,\ldots,h_t-i-1\}
\]
where $\pmb{\xi}^r_{h_t}$, $\pmb{\xi}^r_{h_t-j}$ and $\pmb{\xi}^r_{h_t-j-i}$ are wind power observations at time periods $h_t$, $h_t-j$ and $h_t-j-i$, respectively. 
The deterministic persistence model simply takes the previous historical observation as the point estimator, i.e., $\pmb{\xi}_{h_t+i} = \pmb{\xi}^r_{h_t}$.

Since the Bayesian nonparametric forecast proposed in \cite{Xie_etal_2018} and persistent models focus on the short term prediction, we consider the unit commitment problem for the intra-day market; see \cite{Analui_Scaglione_2017}. The one-hour ahead intra-day unit commitment problem has the planning horizon with $n_h=4$ hours and we make the unit commitment decision of the $t$-th hour at hour $h_t$, where
$
h_t \equiv (\ceil{\frac{t}{n_h} } -1)\cdot n_h.
$ It means that we make the one-hour ahead intra-day unit commitment decisions every four hours. For example, we make $t=1, \ldots, 4$ hour intra-day unit commitment decisions at $h_t=0$.
Suppose that all three generators in the six-bus system are fast start generators that can be committed/de-committed at the intra-day market. 

At any time period $h_t+1$, we can use $m$ latest historical data for the wind power prediction, $\mathcal{D}=\{\xi^r_{h_t-m+1}, \ldots, \xi^r_{h_t} \}$, where $\xi^r_{h_t-i}$ is the hourly wind power observation happened $i$ hours prior to $h_t$. Following  \cite{Xie_etal_2018}, we set $m=100$. For the Bayesian nonparametric forecast approach, we apply the sampling procedure developed in \cite{Xie_etal_2018}. For the probabilistic persistence model, we apply the sampling procedure developed in \cite{Gneiting_etal_2007}. 
%Specifically, the empirical  predictive distribution  for $\pmb{\xi}_{h_t+i}$ specified by
% $
% \{\pmb{\xi}_{h_t}-\pmb{\xi}_{h_t-j}+\pmb{\xi}_{h_t-j-i}: j=0,\ldots,h_t-i-1\}.
% $	}

%When we compare the performance of these approaches, since $F^c$ is unknown, we use the \textit{real} dispatch costs for evaluation \cite{Zhou_etal_2013}. 
Denote the optimal unit commitment decisions between 
hours $h_t+1$ and $h_t+n_h$ on day $d$ obtained from data-driven SUC, empirical SUC and deterministic UC by  $\widehat{\mathbf{u}}^{\star, p}_{d, h_t}$, $\widehat{\mathbf{u}}^{\star e}_{d, h_t}$ and $\widehat{\mathbf{u}}^{\star d}_{d, h_t}$ for $d= 1 , \ldots, n_d$ and $h_t =0, 4, \ldots, 20$. Let the set $\mathcal{H}\equiv\{0, 4, \ldots, 20\}$.
Since $F^c$ is unknown, we evaluate the performance of these solutions by the \textit{actual} incurred cost; see the details in \cite{Zhou_etal_2013}. Denote $\pmb{\xi}_{d,h_t}^r \equiv ({\xi}^r_{d,(h_t+1)}, \ldots, {\xi}^r_{d,(h_t+ n_h)})$ as the real wind power realizations between hours $h_t+1$ and $h_t+n_h$ on day $d$. Then, the real cost $G^r_{d,h_t}$ including both commitment and economic dispatch costs is,
\[
G^r_{d, h_t}(\mathbf{u}_{d, h_t}) \equiv  \mathbf{C}^{su} \mathbf{u}_{d,h_t} + \min_{\mathbf{y}} \mathbf{C}^{fuel} \mathbf{y}(\mathbf{u}_{d,h_t}, \pmb{\xi}_{d,h_t}^{r}). 
\] 
Thus, the total cost occurring in the $n_d$ days is
\begin{equation}
\sum G^\gamma_r \equiv \sum_{d=1}^{n_d} \sum_{h_t\in\mathcal{H}} G^r_{d,h_t}( \widehat{\mathbf{u}}^{\star, \gamma}_{d,h_t} )
\label{eq.G_r}
\end{equation}
where $\gamma=p$ is for the data-driven SUC with the Bayesian nonparametric forecast model \cite{Xie_etal_2018}, $\gamma=e$ is for the empirical SUC with the probabilistic persistent model, and $\gamma=d$ is for the deterministic UC with the persistent model.

The wind power data of 2006 October are used to evaluate the performance of these approaches. 
The results are recorded in Table~\ref{table: sixbusoct}.
We also report the relative savings with respect to  deterministic UC, 
	\[
	r\Delta G = \frac{\sum G^d_r-\sum G^p_r}{\sum G^d_r} ~~~ \mbox{and} ~~~
	r\Delta G = \frac{\sum G^d_r-\sum G^e_r}{\sum G^d_r},
	\]
obtained by the data-driven and empirical SUC approaches.
 The data-driven SUC leads to the total aggregated cost $2,822,705$, %which equals to $91,055$ dollars per day on average, 
while the empirical SUC has a total incurred cost $2,969,178$. %, which corresponds to $95,780$ dollars per day on average. 
Thus, our proposed method has a total saving $146,473$, which represents a $5\%$ savings. Compared with the deterministic UC, the advantage of our data-driven SUC  is even more substantial and it leads to a total saving $638,756$. That means our data-driven SUC could provide better and more reliable performance. 
\begin{table}[]
	\centering
	\caption{Total costs obtained by data-driven SUC, empirical SUC, and deterministic UC approaches when there is no strong prior knowledge on wind power generation input model}
	\label{table: sixbusoct}
	\scalebox{1.1}{
{
			\begin{tabular}{|l|l|l|}
				\hline
				& Total Cost &  $r\Delta G$\\  \hline
				Data-Driven SUC: 	$\sum G^p_r$ &      2,822,705     &  18.4\%     \\ \hline
				Empirical SUC:	$\sum G^e_r$  &     2,969,178 &   14.2\%  \\ \hline
				Deterministic UC: 	$\sum G^d_r$  &     3,461,462 & 0    \\ \hline 
		\end{tabular}}
	}
\end{table}

\textit{Therefore, since the data-driven SUC rigorously considers the wind power generation inherent stochastic uncertainty and input model estimation uncertainty, it outperforms the empirical SUC and deterministic UC.}

%The results also show the robustness and superiority  of our models, which  constantly produces promising unit commitment decisions under various settings.  
%Next, we present analysis for selected days with distinguish wind supply conditions. We first present the analysis for Oct 30th,  which has high wind energy supply. Throughout this day, the wind energy can support $52\%$ total load request on average. Both models provide accurate wind power prediction and perform reasonably well for this day. Our model has an incurred cost $44,259$ while the persistence model has a total cost $40,189$. This indicates both models are suitable for the days with high wind supplies, while the persistence may have an edge here. 

%Finally, we consider the analysis for Oct 24th, which has low wind energy supply, where the wind energy can support merely $10\%$ total load request on average. Our model produces a unit commitment decision with a total incurred cost $116,250$, while the unit commitment decision provided by the persistence model has a total incurred cost $144,983$. Thus, our model has a better prediction performance. 

\vspace{-0.in}

\subsubsection{Performance of OPSEL Approach}
\label{subsubsec:procedurePerformance}

\begin{figure}[h!]
	\vspace{0.in}
	{
		\centering
		\includegraphics[scale=0.3]{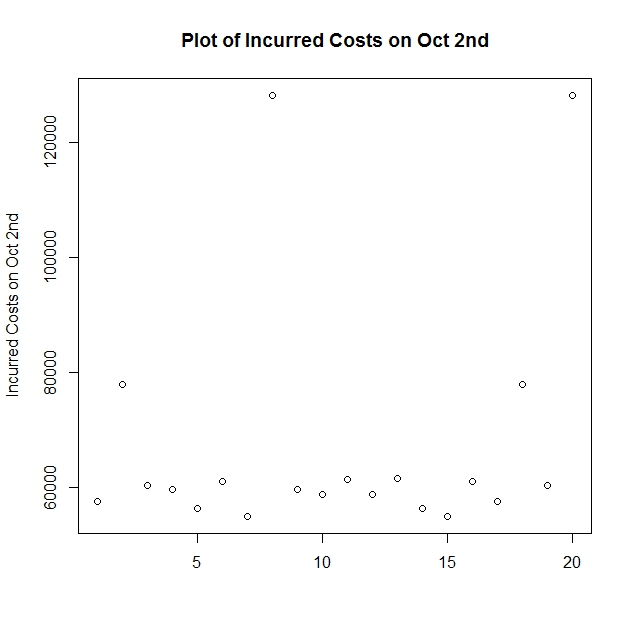}
		\vspace{-0.2in}
		\caption{The scatter plot of incurred costs by using the optimal UC decisions obtained from $L=20$ CPUs to illustrate the impact of finite sampling uncertainty from SAA.}
		\label{fig:Figure2}
	}
\end{figure}

In this section, we use the intra-day unit commitment problem described in Section~\ref{subsubsec: sixpaIMSAR}
to study the performance of the OPSEL approach developed to hedge against the finite sampling uncertainty induced by using SAA. We consider the data-driven SUC with the nonparametric Bayesian forecast model \cite{Xie_etal_2018}.
To illustrate the impact of finite sampling uncertainty, we use a representative day, October 2nd, for demonstration. Fig.~\ref{fig:Figure2} plots the incurred cost $\sum_{h_t\in \mathcal{H}} G^r_{d, h_t}(\widehat{\mathbf{u}}^{\star, p}_{d,h_t,\ell})$ for $\ell=1,\ldots,L$ obtained by utilizing $L=20$ CPUs to independently solve the SAA approximated data-driven SUC optimization problem in (\ref{eq: genapp}), where $\widehat{\mathbf{u}}^{\star, p}_{d,h_t,\ell}$ is  obtained from the $\ell$-th CPU. Each decision is obtained by using SAA with the number of scenarios, $S=50$. In the plot, each dot represents one optimal decision, the horizontal axis provides the CPU index $\ell$, and the vertical axis provides the incurred cost. We can observe that the cost is widespread, ranging from about $55,000$ to as much as around $120,000$. \textit{Thus, Fig.~\ref{fig:Figure2} demonstrates the obvious impact of finite sampling error induced by SAA.} %Since it is hard to know which scenario set could lead to promising results, it is necessary to exploit the parallel computing which can generate multiple SAA optimal unit commitment candidate decisions and select the best decision from them. 

\vspace{-0.in}

Then, we study the performance of OPSEL approach introduced in Section~\ref{subsec: OCBA}.  Suppose there are $L$ CPUs available for parallel computing. Specifically, for each intra-day unit commitment problem at the $h_t$-th hour on day $d$, we consider the sample average approximated data-driven SUC in (\ref{eq: genapp}) with $S=50$. By solving $L$ intra-day unit commitment problems in parallel, we obtain the optimal unit commitment decisions for the next $n_h$ hours, denoted by $\widehat{\mathbf{u}}^{\star, p}_{d, h_t, 1}, \ldots, \widehat{\mathbf{u}}^{\star, p}_{d, h_t, L}$. 
After that, the OPSEL is used to estimate the objective values $G^p (\widehat{\mathbf{u}}^{\star,p}_{d,h_t, 1} ), \ldots, G^p (\widehat{\mathbf{u}}^{\star,p}_{d,h_t, L} )$ and efficiently select the best unit commitment decision, denoted by  $\widehat{\mathbf{u}}^{\star, p}_{d,h_t, b}$.

We compare the proposed OPSEL with the sample average approximated data-driven SUC approach that ignores the finite sampling error by using the wind power data occurring in October $1$st--$31$st. We evaluate these two approaches with the \textit{actual} incurred cost. 
If we ignore the finite sampling error, the total cost is $\sum G^p_r=\sum_{d=1}^{n_d} \sum_{h_t \in \mathcal{H}} G^r_{d,h_t}( \widehat{\mathbf{u}}^{\star, p}_{d,h_t} )$.  For the OPSEL approach, the total incurred cost is
\[
\sum G^{p,b}_r \equiv \sum_{d=1}^{n_d} \sum_{h_t\in\mathcal{H}} G^r_{d,h_t}(\widehat{\mathbf{u}}^{\star, p}_{d,h_t, b}),
\] 
where $
G^r_{d,h_t}(\mathbf{u}_{d, h_t, b}) \equiv  \mathbf{C}^{su}\widehat{\mathbf{u}}^{\star,p}_{d,h_t, b} + \min_{\mathbf{y}} \mathbf{C}^{fuel} \mathbf{y}(\widehat{\mathbf{u}}^{\star,p}_{d,h_t, b}, \pmb{\xi}_{d,h_t}^{r}) 
$ and it is the cost occurring between hours $h_t+1$ and $h_t+n_h$ on day $d$.

% \[
%G^r_{d, h_t}(\widehat u_{i, d, h_t}) \equiv  C_1\widehat{\mathbf{u}}^{\star p}_{dh_t, b} + \min_{\mathbf{y}} C_2 \mathbf{y}(\widehat{\mathbf{u}}^{\star p}_{dh_t, b}, \pmb{\xi}_{dh_t}^{r}). 
% \] 

Here, we use $L=4$ CPUs. The results of total operation cost in October obtained by using the data-driven SUC with and without OPSEL are recorded in Table~\ref{table: OPSEL}. The OPSEL approach can control the impact of finite sampling error induced by SAA and leads to a cheaper total cost, $
\sum G^{p,b}_r  \leq \sum	 G^p_r.
$
The relative cost saving caused by using the proposed OPSEL approach,
$
	r\Delta G \equiv \frac{\sum	 G^p_r - \sum	 G^{p,b}_r }{\sum G^p_r},
$	
is about 20\%. To further check the robustness of our approach, we study the performance of OPSEL under various settings of $\Delta T$ and $T$.  It is obvious that our procedure produces substantially better $\sum G^{p,b}_r$ solutions than $\sum	 G^p_r$ under all settings of $\Delta T$ and $T$. \textit{Therefore, our OPSEL approach can effectively control the impact of finite sampling error and significantly improve the reliability and cost-efficiency of unit commitment decisions.} 
%In particular, our OPSEL procedure brings $535,452$ or $19.0\%$ cost savings over the classical L-Shaped algorithm.  Thus, by controlling the impact from SAA finite sampling estimation error, our optimization procedure can lead to better and more robust unit commitment decision.

\begin{table}[]
	\centering
	\caption{The results of total incurred costs obtained by data-driven SUC with and without the proposed OPSEL}
	\label{table: OPSEL}
	\scalebox{0.96}{
	{
			\begin{tabular}{|l|l|l|}
				\hline
				& Cost        &         $r\Delta G$           \\ \hline
				Data-driven SUC without OPSEL $\sum G^p_r$  &    2,822,705      & 0              \\ \hline
				SUC + OPSEL $\sum G^{p,b}_r $ $(T=500,\Delta T=100)$ &  2,323,507 & 17.7\% \\ \hline
				SUC + OPSEL $\sum G^{p,b}_r $ $(T=1000,\Delta T=50)$ &   2,186,869 & 22.5\% \\ \hline
				SUC + OPSEL $\sum G^{p,b}_r $ $(T=1000,\Delta T=100)$ & 2,282,531  &  19.1\% \\ \hline
				SUC + OPSEL $\sum G^{p,b}_r $ $(T=1000,\Delta T=200)$ &  2,134,574	& 24.4\%\\ \hline
				%OPSEL approach $\sum G^{p,b}_r $ $(T=2000,\Delta T=100)$ &  2,219,266 & 21.4\% \\ \hline
		\end{tabular}}
	}
\end{table}

We also record the overhead computational burden induced by the OPSEL candidate selection. We observe that the total running time of first-stage optimization search by using the L-Shaped optimization for the one-month period is about $40,000$ seconds, while the total time spent on the candidate selection is around $1,500$ seconds. Thus, the overhead burden is negligible. 
%Given the substantial improvements recorded in the proposed OPSEL method, our method is suprior and can be widely implemented.

\subsection{Empirical Study with 118-Bus System}
\label{subsec:118Bus}

To evaluate the scalability and robustness of our approach, in this section, we consider the derivative 118-bus system including 54 thermal units, three wind farms and 186 transmission lines; see the description in \cite{pena2017extended}. Similar to the 6-bus system case, we assume deterministic loads and stochastic wind supply\cite{Zhou_etal_2013} and use the 2006 data of the U.S. Illinois power system for the load and the wind power generation. The whole system's wind penetration $R$ is 9.9\%. 
%Due to the relatively low percentage of wind power generation with respect to total load, we increase the wind power by 3 times to match the same proportion as in 6-bus system. 
In addition, we set the wind curtailment at 30\$/MWh and load shedding price at 3500\$/MWh. We just consider constraints (4)-(9) in \cite{Wang_etal_2017} as in section \ref{subsec:sixBus}.

\subsubsection{Case Study with Parametric Forecast Model}
\label{subsec:118bus_parametric}

One month wind power data are used to study the performance of the proposed data-driven SUC and the empirical SUC. We use the same settings with those used in Section~\ref{subsubsec: sixpacompound}, and consider one day-ahead unit commitment problem here. % with 24 hours planning horizon. 
%Let $m=1$ to consider the situation with very limited amount of historical data.
When we solve the SUC problems for the optimal decision estimates $\widehat{\mathbf{u}}_{d}^{\star, e }$ and $\widehat{\mathbf{u}}_{d}^{\star, p }$, we set the scenario size to be $S=50$. Then, we use $S_e = 1000$ to evaluate the true expected cost $\sum G^p $ and $\sum G^e$. The cases with standard deviation $\phi_{d,t} =5\%\mu_{d,t}^c, 10\%\mu_{d,t}^c, 20\% \mu_{d,t}^c$ and $m = 1, 5, 10$  are analyzed and the results are recorded in Table~\ref{table: 118busr}. %We consider the wind penetration levels $R=9.9\%,39.7\%$.
As $m$ goes big, the input model estimation uncertainty decreases. %, the advantage of proposed approach will be smaller.
\textit{According to Table \ref{table: 118busr}, the advantage of proposed data-driven SUC increases as the wind penetration increases, the wind power generation variation becomes larger, and the amount of valid historical data decreases.} 

\begin{comment}
\begin{table}[h!]
\centering
\caption{Total expected costs when $R=9.9\%,39.7\%$ and $\phi_{d,t} =5\%\mu_{d,t}^c, 10\%\mu_{d,t}^c, 20\% \mu_{d,t}^c$}
\label{table: 118busr}
\begin{tabular}{|c|c|c|c|c|}
\hline
$R$                    &       Standard Deviation                         & $\sum G^p$    & $\sum G^e $   & $r\Delta G$ \\ \hline
\multirow{3}{*}{9.9\%}  & $\phi_{d,t}= 5\% \mu_{d,t}^c$  & 64,701,130.32 & 65,874,626.24 & 1.8\%      \\ \cline{2-5} 
                         & $\phi_{d,t}= 10\% \mu_{d,t}^c$ & 65,609,752.49 & 67,157,709.28 & 2.4\%      \\ \cline{2-5} 
                         & $\phi_{d,t}= 20\% \mu_{d,t}^c$ & 65,065,808.32 & 69,475,458.36 & 6.8\%      \\ \hline
\multirow{3}{*}{39.7\%} & $\phi_{d,t}= 5\% \mu_{d,t}^c$  & 53,555,389.28 & 59,608,154.35 & 11.3\%     \\ \cline{2-5} 
                         & $\phi_{d,t}= 10\% \mu_{d,t}^c$ & 50,857,923.59 & 62,740,670.89 & 23.4\%     \\ \cline{2-5} 
                         & $\phi_{d,t}= 20\% \mu_{d,t}^c$ & 48,772,789.95 & 66,963,228.08 & 37.3\%     \\ \hline
\end{tabular}
\end{table}
\end{comment}

\begin{table}[h!]
\centering
\caption{Total expected costs when $\phi_{d,t} =5\%\mu_{d,t}^c, 10\%\mu_{d,t}^c, 20\% \mu_{d,t}^c$ and $m = 1, 5, 10$}
\label{table: 118busr}
\begin{tabular}{|c|c|c|c|c|}
\hline
Standard Deviation             & $m = 1$ & $m = 5$ & $m = 10$ \\ \hline
$\phi_{d,t}= 5\% \mu_{d,t}^c$  & 8.22\%  & 1.26\%      & 0.11\%   \\ \hline
$\phi_{d,t}= 10\% \mu_{d,t}^c$ & 13.80\% & 5.51\%  & 0.14\%   \\ \hline
$\phi_{d,t}= 20\% \mu_{d,t}^c$ & 20.59\% & 12.36\%  & 2.60\%    \\ \hline
\end{tabular}
\end{table}

%when $\phi_{d,t}= 20\% \mu_{d,t}^c$, our approach can outperforms the empirical approach by 6.78\%. 
%\textit{The results show that our data-driven SUC still haves significant effect when implemented upon larger system %with high renewable penetration for parametric cases.}

\vspace{-0.1in}

\subsubsection{Case study with Nonparametric Forecast Model}
\label{subsec:118bus_nonparametric}

Here we consider the case when there is no strong prior information on the underlying wind power generation distribution and the nonparametric distributions are used for wind power forecast, which is similar %to 
with that used in Section~\ref{subsubsec: sixpaIMSAR}. We use the 118-bus test case to study the performance of: (1) deterministic unit commitment, (2) empirical SUC and (3) data-driven SUC. 
We utilize the deterministic persistence as prediction model for the deterministic unit commitment, use the probabilistic persistence model for the empirical SUC, and implement Bayesian nonparametric wind power forecast model \cite{Xie_etal_2018} for data-driven SUC with prediction risk accounting for both wind generation stochastic uncertainty and model estimation error. 
We consider the unit commitment problem for the intra-day market with 4 hours planning horizon and set the amount of historical data used for the forecast to be $m = 100$. %\textcolor{red}{(which part of wind power data is used?)} 
One month wind power data are used to evaluate the performance and the results are recorded in Table \ref{table: 118busoct}.
%When $R=9.9\%$, 
The empirical SUC, accounting for wind power generation stochastic uncertainty only, leads to 1.12\% savings compared with the deterministic UC method. The proposed data-driven SUC, accounting for both wind power stochastic uncertainty and input model estimation uncertainty, leads to 2.58\% savings.
\textit{Thus, the proposed data-driven SUC, accounting for all sources of uncertainties, can lead to more cost-efficient and robust decisions, and the advantage is larger as the wind power penetration increases.}

%\begin{table}[h!]
%\centering
%\caption{Total costs of operation obtained from various approaches}
%\label{table: 118busoct}
%\begin{tabular}{|l|l|l|l|}
%\hline
%$R$                    &        Approach                        & Total Cost     & $r\Delta G$ \\ \hline
%\multirow{3}{*}{9.9\%}  & Data-Driven SUC: $\sum G^p_r$  & 51,754,667.24  & 36.7\%     \\ \cline{2-4} 
%                         & Empirical SUC: $\sum G^e_r$    & 59,918,021.64  & 18.1\%     \\ \cline{2-4} 
%                         & Deterministic UC: $\sum G^d_r$ & 70,741,233.73  & 0           \\ \hline
%\multirow{3}{*}{39.7\%} & Data-Driven SUC: $\sum G^p_r$  & 157,186,751.57 & 68.8\%     \\ \cline{2-4} 
%                         & Empirical SUC: $\sum G^e_r$    & 202,139,110.11 & 31.3\%     \\ \cline{2-4} 
%                         & Deterministic UC: $\sum G^d_r$ & 265,399,353.59 & 0           \\ \hline
%\end{tabular}
%\end{table}

\begin{table}[h!]
\centering
\caption{Total costs of operation obtained from various approaches when $R=9.9\%$}
\label{table: 118busoct}
\begin{tabular}{|l|l|l|l|l|}
\hline
Approach                       & Penalty Cost & Penalty Ratio & Total Cost    & $r\Delta G$ \\ \hline
Deterministic UC: $\sum G^d_r$ &  1,156,456.31 & 2.21\%       &  52,327,089.47 & 0.00\%      \\ \hline
Empirical SUC: $\sum G^e_r$    &  537,661.31  & 1.04\%        &  51,743,500.01 & 1.12\%      \\ \hline
Data-Driven SUC: $\sum G^p_r$  & 0.00         & 0.00\%        &  50,974,697.80 & 2.58\%      \\ \hline
\end{tabular}
\end{table}

We also record the \textit{penalty cost} induced when the energy production does not meet the demand; see Eq.~(\ref{eq: sixbusobj}). The proportions of penalty cost obtained by the deterministic unit commitment, the empirical SUC, and the proposed data-driven SUC are shown in Table \ref{table: 118busoct}. The decision made by the deterministic UC method leads to %11 cases over 180 planning horizons have demand unsatisfied problem with 
the total penalty cost \$  1,156,456.31. The proportion of the penalty cost to the total cost $ \sum G^d_r $ is 2.21\%. 
The empirical SUC method
%, the number of issue happened is 5 with
has the total penalty cost \$ 537,661.31 contributing 1.04\% percentage of the total cost  $ \sum G^e_r $. 
\textit{Thus, the results indicate that the load unmet risk can be hedged by utilizing the proposed data-driven SUC, % at both frequency and magnitude aspects. 
and it can lead to cost saving, more reliable and robust power grids with high renewable energy penetration.} %The merit of hedging demand unsatisfied risk induced by Data-Driven SUC explains and confirms the best performance of total cost saving advantage.

\subsubsection{Performance of  OPSEL Approach}
\label{subsubsec:procedurePerformance118}
Similar to Section~\ref{subsubsec:procedurePerformance}, we investigate the performance of the proposed OPSEL approach by using the 118-bus system. Let $L=3$, $\Delta T = 50$ and $T = 500$. The results of total operation cost obtained by using the data-driven SUC with and without OPSEL are recorded in Table~\ref{table: OPSEL118}. The relative cost saving by implementing our OPSEL approach is about 2.36\%. \textit{Thus, the proposed OPSEL approach demonstrates the promising performance on the large-scale system by hedging against the impact of finite sampling error induced by SAA.}

\begin{table}[]
	\centering
	\caption{Total incurred costs of operation on 118-bus system}
	\label{table: OPSEL118}
	\scalebox{0.96}{
	{
			\begin{tabular}{|l|l|l|}
				\hline
				& Cost        &         $r\Delta G$           \\ \hline
				Data-driven SUC without OPSEL $\sum G^p_r$  &   50,974,697.80       & 0.00\%              \\ \hline
				SUC + OPSEL $\sum G^{p,b}_r $ $(T=50,\Delta T=500)$ &   49,771,552.75 & 2.36\% \\ \hline
		\end{tabular}}
	}
\end{table}

\section{CONCLUSION}
\label{sec: conclusion}  	
In this paper, we first propose a data-driven stochastic optimization to guide the power system unit commitment decision, which can simultaneously hedge against the underlying stochastic uncertainty or volatility of wind power generation and statistical input model estimation error. Then, we introduce an optimization and selection approach that can efficiently utilize the parallel computing to quickly find the optimal unit commitment decision, while controlling the impact of finite sampling error induced by SAA. Various case studies on a six-bus system and a 118-bus system verify the advantages of our proposed data-drive SUC for both parametric and nonparametric forecast models. %\textcolor{red}{And scalability are assessed on 118-bus system.} 
They also demonstrate that our OPSEL procedure can further deliver the optimal unit commitment decision hedging against the impact of finite sampling error.

%\begin{acknowledgements}
%\end{acknowledgements}




\bibliographystyle{unsrt} 
\bibliography{grid}

\begin{thebibliography}{10}

\bibitem{Hargreaves_Hobbs_2012}
J.~J. Hargreaves and B.~F. Hobbs.
\newblock Commitment and dispatch with uncertain wind generation by dynamic
  programming.
\newblock {\em IEEE Transactions on Sustainable Energy}, 3(4):724 -- 734, 2012.

\bibitem{Zhou_etal_2013}
Z.~Zhou, A.~Botterud, J.~Wang, R.J. Bessa, H.~Keko, J.~Sumaili, and V.~Miranda.
\newblock Application of probabilistic wind power forecasting in electricity
  markets.
\newblock {\em Wind Energy}, 16(3):321--338, 2013.

\bibitem{Zhao_Wu_2014}
Z.~Zhao and L.~Wu.
\newblock Impacts of high penetration wind generation and demand response on
  lmps in day-ahead market.
\newblock {\em IEEE Transaction On Smart Grid}, 5(1):220--229, 2014.

\bibitem{Jiang_Wang_Guan_2012}
R.~Jiang, J.~Wang, and Y.~Guan.
\newblock Robust unit commitment with wind power and pumped storage hydro.
\newblock {\em IEEE Transactions on Power Systems}, 27(2):800--810, 2012.

\bibitem{Papavasiliou_Oren_2013}
A.~Papavasiliou and S.~S. Oren.
\newblock Multiarea stochastic unit commitment for high wind penetration in a
  transmission constrained network.
\newblock {\em Operations Research}, 61(3):578--592, 2013.

\bibitem{Wang_Shahidehpour_Li_2008}
J.~Wang, M.~Shahidehpour, and Z.~Li.
\newblock Security-constrained unit commitment with volatile wind power
  generation.
\newblock {\em IEEE Transactions on Power Systems}, 23(3):1319 -- 1327, 2008.

\bibitem{Ruiz_etal_2009}
P.~A. Ruiz, C.~R. Philbrick, and P.~W. Sauer.
\newblock Wind power day-ahead uncertainty management through stochastic unit
  commitment policies.
\newblock In {\em 2009 IEEE/PES Power Systems Conference and Exposition}, pages
  1--9, 2009.

\bibitem{Tuohy_etal_2009}
A.~Tuohy, P.~Meibom, E.~Denny, and M.~O'Malley.
\newblock Unit commitment for systems with significant wind penetration.
\newblock {\em IEEE Transactions on Power Systems}, 24(2):592 -- 601, 2009.

\bibitem{Wang_Guan_Wang_2012}
Q.~Wang, Y.~Guan, and J.~Wang.
\newblock A chance-constrained two-stage stochastic program for unit commitment
  with uncertain wind power output.
\newblock {\em IEEE Transactions on Power Systems}, 27(1):592 -- 601, 2012.

\bibitem{Xie_etal_2018}
W.~Xie, P.~Zhang, R.~Chen, and Z.~Zhou.
\newblock A nonparametric bayesian framework for short-term wind power
  probabilistic forecast.
\newblock {\em Accepted}, 2018.

\bibitem{Takriti_etal_1996}
S.~Takriti, J.~Birge, and E.~Long.
\newblock A stochastic model for the unit commitment problem.
\newblock {\em IEEE Transactions on Power Systems}, 11(3):1497--1508, 1996.

\bibitem{madaeni2013impacts}
Seyed~Hossein Madaeni and Ramteen Sioshansi.
\newblock The impacts of stochastic programming and demand response on wind
  integration.
\newblock {\em Energy Systems}, 4(2):109--124, 2013.

\bibitem{Huang_Zheng_Wang_2014}
Y.~Huang, Q.~P. Zheng, and J.~Wang.
\newblock Two-stage stochastic unit commitment model including non-generation
  resources with conditional value-at-risk constraints.
\newblock {\em Electric Power Systems Research}, 116:427--438, 2014.

\bibitem{Bertsimas_etal_2013}
D.~Bertsimas, E.~Litvinov, X.~A. Sun, J.~Zhao, and T.~Zheng.
\newblock Adaptive robust optimization for the security constrained unit
  commitment problem.
\newblock {\em IEEE Transactions on Power Systems}, 28(1):52 -- 63, 2013.

\bibitem{Zhao_etal_2013}
C.~Zhao, J.~Wang, and J.-P. Watson.
\newblock Multi-stage robust unit commitment considering wind and demand
  response uncertainties.
\newblock {\em IEEE Transactions on Power Systems}, 28(3):2708 -- 2718, 2013.

\bibitem{Lee_etal_2014}
C.~Lee, C.~Liu, S.~Mehrotra, and M.~Shahidehpour.
\newblock Modeling transmission line constraints in two-stage robust unit
  commitment problem.
\newblock {\em IEEE Transactions on Power Systems}, 29(3):1221 -- 1231, 2014.

\bibitem{Xiong_etal_2017}
P.~Xiong, P.~Jirutitijaroen, and C.~Singh.
\newblock A distributionally robust optimization model for unit commitment
  considering uncertain wind power generation.
\newblock {\em IEEE Transactions on Power Systems}, 32(1):39 -- 49, 2017.

\bibitem{Zhao_Zeng_2012}
L.~Zhao and B.~Zeng.
\newblock Robust unit commitment problem with demand response and wind energy.
\newblock In {\em Power and Energy Society General Meeting}, pages 1--8. IEEE,
  2012.

\bibitem{Lorca_Sun_2015}
\'{A}. Lorca and X.~A. Sun.
\newblock Adaptive robust optimization with dynamic uncertainty sets for
  multi-period economic dispatch under significant wind.
\newblock {\em IEEE Transactions on Power Systems}, 30(4):1702 -- 1713, 2015.

\bibitem{kazemzadeh2019robust}
Narges Kazemzadeh, Sarah~M Ryan, and Mahdi Hamzeei.
\newblock Robust optimization vs. stochastic programming incorporating risk
  measures for unit commitment with uncertain variable renewable generation.
\newblock {\em Energy Systems}, 10(3):517--541, 2019.

\bibitem{Bian_etal_2015}
Q.~Bian, H.~Xin, Z.~Wang, and K.~P. Wong.
\newblock Distributionally robust solution to the reserve scheduling problem
  with partial information of wind power.
\newblock {\em IEEE Transactions on Power Systems}, 30(5):2822--2823, 2015.

\bibitem{Wang_etal_2016J}
Z.~Wang, Q.~Bian, H.~Xin, and D.~Gan.
\newblock A distributionally robust co-ordinated reserve scheduling model
  considering cvar-based wind power reserve requirements.
\newblock {\em IEEE Transactions on Sustainable Energy}, 7(2):625--636, 2016.

\bibitem{Jiang_etal_2013}
R.~Jiang, J.~Wang, M.~Zhang, and Y.~Guan.
\newblock Two-stage minimax regret robust unit commitment.
\newblock {\em IEEE Transactions on Power Systems}, 28(3):2271 -- 2282, 2013.

\bibitem{ozturk2003stochastic}
Ugur~Aytun Ozturk.
\newblock {\em The stochastic unit commitment problem: a chance constrained
  programming approach considering extreme multivariate tail probabilities}.
\newblock PhD thesis, University of Pittsburgh, 2003.

\bibitem{ozturk2004solution}
U~Aytun Ozturk, Mainak Mazumdar, and Bryan~A Norman.
\newblock A solution to the stochastic unit commitment problem using chance
  constrained programming.
\newblock {\em IEEE Transactions on Power Systems}, 19(3):1589--1598, 2004.

\bibitem{wang2011chance}
Qianfan Wang, Yongpei Guan, and Jianhui Wang.
\newblock A chance-constrained two-stage stochastic program for unit commitment
  with uncertain wind power output.
\newblock {\em IEEE Transactions on Power Systems}, 27(1):206--215, 2011.

\bibitem{shapiro_2009}
A.~Shapiro, D.~Dentcheva, and A.~Ruszczynski.
\newblock {\em Lectures on Stochastic Programming: Modeling and Theory}.
\newblock SIAM, Philadelphia, 2009.

\bibitem{Chen_etal_2000}
C.~H. Chen, J.~Lin, E.~Y{u}cesan, and S.~E. Chick.
\newblock Simulation budget allocation for further enhancing the efficiency of
  ordinal optimization.
\newblock {\em Journal of Discrete Event Dynamic Systems: Theory and
  Applications}, 10:251--270, 2000.

\bibitem{Boesel_Nelson_Kim_2003}
J.~Boesel, B.~L. Nelson, and S.-H. Kim.
\newblock Using ranking and selection to “clean up” after simulation
  optimization.
\newblock {\em Operations Research}, 51(5):814 -- 825, 2003.

\bibitem{Kim_Nelson_2007}
Seong-Hee Kim and B.~L. Nelson.
\newblock Recent advances in ranking and selection.
\newblock In {\em 2007 Winter Simulation Conference}, pages 162--172, Dec 2007.

\bibitem{Powell_Ryzhov_2012}
Warren~B. Powell and Ilya~O. Ryzhov.
\newblock {\em Ranking and Selection}, chapter~4, pages 71--88.
\newblock Wiley-Blackwell, 2012.

\bibitem{Kim_Nelson_2001}
Seong-Hee Kim and Barry~L. Nelson.
\newblock A fully sequential procedure for indifference-zone selection in
  simulation.
\newblock {\em ACM Transactions on Modeling and Computer Simulation},
  11(3):251--273, 2001.

\bibitem{Chick_Inoue_2001}
Stephen~E. Chick and Koichiro Inoue.
\newblock New two-stage and sequential procedures for selecting the best
  simulated system.
\newblock {\em Operations Research}, 49(5):732--743, 2001.

\bibitem{Frazier_Powell_Dayanik_2008}
Peter~I. Frazier, Warren~B. Powell, and Savas Dayanik.
\newblock A knowledge-gradient policy for sequential information collection.
\newblock {\em SIAM Journal on Control and Optimization}, 47(5):2410--2439,
  2008.

\bibitem{Chick_Branke_Schmidt_2010}
S.~E. Chick, J.~Branke, and C.~Schmidt.
\newblock Sequential sampling to myopically maximize the expected value of
  information.
\newblock {\em INFORMS Journal on Computing}, 22(1):71--80, 2010.

\bibitem{Chen_etal_2008}
Chun-Hung Chen, Donghai He, Michael Fu, and Loo~Hay Lee.
\newblock Efficient simulation budget allocation for selecting an optimal
  subset.
\newblock {\em INFORMS Journal on Computing}, 20(4):579--595, 2008.

\bibitem{xiao2013optimal}
Hui Xiao, Loo~Hay Lee, and Kien~Ming Ng.
\newblock Optimal computing budget allocation for complete ranking.
\newblock {\em IEEE Transactions on Automation Science and Engineering},
  11(2):516--524, 2013.

\bibitem{xu2015simulation}
Jie Xu, Edward Huang, Chun-Hung Chen, and Loo~Hay Lee.
\newblock Simulation optimization: A review and exploration in the new era of
  cloud computing and big data.
\newblock {\em Asia-Pacific Journal of Operational Research}, 32(03):1550019,
  2015.

\bibitem{Glynn_Juneja_2004}
P.~Glynn and S.~Juneja.
\newblock A large deviations perspective on ordinal optimization.
\newblock In R.~G. Ingalls, M.~D. Rossetti, J.~S. Smith, and B.~A. Peters,
  editors, {\em Proceedings of the 2004 Winter Simulation Conference}, pages
  101--112. IEEE Computer Society, Washington, DC, 2004.

\bibitem{Ryzhov_2016}
I.~O. Ryzhov.
\newblock On the convergence rates of expected improvement methods.
\newblock {\em Operations Research}, 64(6):1515 -- 1528, 2016.

\bibitem{Quan_etal_2013}
N.~Quan, J.~Yin, S.~H. Ng, and L.~H. Lee.
\newblock Simulation optimization via kriging: a sequential search using
  expected improvement with computing budget constraints.
\newblock {\em IIE Transactions}, 45(7):763--780, 2013.

\bibitem{Zheng_Wang_Liu_2015}
Q.~P. Zheng, J.~Wang, and A.~L. Liu.
\newblock Stochastic optimization for unit commitment - a review.
\newblock {\em IEEE Transactions on Power Systems}, 30(4):1913 -- 1924, 2015.

\bibitem{Wang_etal_2016}
Yishen Wang, Zhi Zhou, Cong Liu, and A.~Botterud.
\newblock Evaluating stochastic methods in power system operations with wind
  power.
\newblock In {\em 2016 IEEE International Energy Conference (ENERGYCON)}, pages
  1--6, 2016.

\bibitem{Zheng_etal_2013}
Qipeng~P. Zheng, Jianhui Wang, Panos~M. Pardalos, and Yongpei Guan.
\newblock A decomposition approach to the two-stage stochastic unit commitment
  problem.
\newblock {\em Annals of Operations Research}, 210(1):387--410, Nov 2013.

\bibitem{Hobbs_Rothkopf_O'Neill__Chao_2001}
B.F. Hobbs, M.H. Rothkopf, R.P. O'Neill, and Hung po~Chao.
\newblock {\em The Next Generation of Electric Power Unit Commitment Models}.
\newblock Kluwer Academic, Norwell, MA, USA, 2001.

\bibitem{zhou2016stochastic}
Z~Zhou, C~Liu, and A~Botterud.
\newblock Stochastic methods applied to power system operations with renewable
  energy: A review.
\newblock Technical report, Argonne National Lab.(ANL), Argonne, IL (United
  States), 2016.

\bibitem{ummels2007impacts}
Bart~C Ummels, Madeleine Gibescu, Engbert Pelgrum, Wil~L Kling, and Arno~J
  Brand.
\newblock Impacts of wind power on thermal generation unit commitment and
  dispatch.
\newblock {\em IEEE Transactions on energy conversion}, 22(1):44--51, 2007.

\bibitem{delarue2008adaptive}
Erik Delarue and William D’haeseleer.
\newblock Adaptive mixed-integer programming unit commitment strategy for
  determining the value of forecasting.
\newblock {\em Applied Energy}, 85(4):171--181, 2008.

\bibitem{Nielsen_etal_1998}
Nielsen~Torben Skov, Joensen Alfred, Madsen Henrik, Landberg Lars, and Giebel
  Gregor.
\newblock A new reference for wind power forecasting.
\newblock {\em Wind Energy}, 1(1):29--34, 1998.

\bibitem{Gneiting_etal_2007}
Gneiting Tilmann, Balabdaoui Fadoua, and Raftery~Adrian E.
\newblock Probabilistic forecasts, calibration and sharpness.
\newblock {\em Journal of the Royal Statistical Society: Series B (Statistical
  Methodology)}, 69(2):243--268, 2007.

\bibitem{Kavasseri_2009}
Rajesh~G. Kavasseri and Krithika Seetharaman.
\newblock Day-ahead wind speed forecasting using f-arima models.
\newblock {\em Renewable Energy}, 34(5):1388 -- 1393, 2009.

\bibitem{Zhang_Wang_Wang_2014}
Yao Zhang, Jianxue Wang, and Xifan Wang.
\newblock Review on probabilistic forecasting of wind power generation.
\newblock {\em Renewable and Sustainable Energy Reviews}, 32:255 -- 270, 2014.

\bibitem{Pandzic_etal_2016}
H.~Pandžić, Y.~Dvorkin, T.~Qiu, Y.~Wang, and D.~S. Kirschen.
\newblock Toward cost-efficient and reliable unit commitment under uncertainty.
\newblock {\em IEEE Transactions on Power Systems}, 31(2):970--982, 2016.

\bibitem{Wang_etal_2017}
Y.~Wang, Z.~Zhou, A.~Botterud, and K.~Zhang.
\newblock Optimal wind power uncertainty intervals for electricity market
  operation.
\newblock {\em IEEE Transactions on Sustainable Energy}, 9(1):199 -- 210, 2018.

\bibitem{pena2017extended}
Ivonne Pena, Carlo~Brancucci Martinez-Anido, and Bri-Mathias Hodge.
\newblock An extended ieee 118-bus test system with high renewable penetration.
\newblock {\em IEEE Transactions on Power Systems}, 33(1):281--289, 2017.

\bibitem{Yang_Berger_1997}
R.~Yang and J.~Berger.
\newblock A catalog of noninformative priors.
\newblock {\em ISDS Discussion Paper}, pages 97--42, 1997.

\bibitem{Analui_Scaglione_2017}
B.~Analui and A.~Scaglione.
\newblock A dynamic multistage stochastic unit commitment formulation for
  intraday markets.
\newblock {\em IEEE Transactions on Power Systems}, 33(4):3653--3663, 2018.

\end{thebibliography}

\end{document}